\documentclass[preprints,article,accept,moreauthors,pdftex]{Definitions/mdpi} 
\pdfoutput=1

\firstpage{1} 
\pubvolume{xx}
\issuenum{1}
\articlenumber{5}
\pubyear{2019}
\copyrightyear{2019}
\history{Received: date; Accepted: date; Published: date}

\usepackage{amssymb}

\newcommand*{\dif}[1]{{\mathrm{d}}#1}  
\newcommand*{\vect}[1]{\mathbf{#1}}  
\newcommand*{\mtx}[1]{\mathbf{#1}}  
\newcommand*{\T}{^T}  
\newcommand*{\true}{^{\mathrm{true}}} 

\DeclareMathOperator{\tr}{tr}

\newcommand*{\doi}[1]{doi: \href{https://doi.org/#1}{\nolinkurl{#1}}}

\Title{Theoretical Limits of Star Sensor Accuracy\texorpdfstring{$\dagger$}{}}

\Author{Marcio A. A. Fialho $^{1,\dagger,}$*\orcidA{} and Daniele Mortari $^{2}$\orcidB{}}

\AuthorNames{Marcio A. A. Fialho and Daniele Mortari}

\address{%
$^{1}$ \quad INPE - National Institute for Space Research, Av. dos Astronautas, 1758. S\~ao Jos\'e dos Campos, Brazil; marcio.fialho@inpe.br\\
$^{2}$ \quad Texas A\&M University, College Station, TX, USA; mortari@tamu.edu}

\corres{Correspondence: marcio.fialho@inpe.br; Tel.: +55-12-3208-6145}

\firstnote{This paper is a revised version of the work previously published in \cite{Fialho2017} as a doctorate thesis chapter.} 

\abstract{To achieve mass, power, and cost reduction, there is a trend to reduce the volume of many instruments aboard spacecraft, especially for small spacecraft (cubesats or nanosats) with very limited mass, volume and power budgets. With the current trend of miniaturizing spacecraft instruments one could naturally ask if is there a physical limit to this process for star sensors. This paper shows that there is a fundamental limit on star sensor accuracy, which depends on stellar distribution, star sensor dimensions and exposure time. An estimate of such limit is given for our location in the galaxy.}


\begin{document}



\section{Introduction}
 
Many progresses in a variety of fields in science and technology could be accomplished thanks to the miniaturization obtained in microelectronics in the recent decades. One of the most remarkable examples is the prediction by Gordon Moore that the computational power would increase exponentially, an empirical observation that became known as Moore's law \cite{Moore, Ball, Charles, Beigel}. Yet, this rate of improvement is not expected to last forever. Eventually a fundamental limit will be reached when the size of transistors reaches atomic scales. Likewise, in other fields of science and technology, fundamental limits to miniaturization and performance improvements are often found. For instance, in the field of telecommunications, there is a theoretical minimum amount of energy that must be spent to transmit a bit in a digital message from one point to another within a given time interval, being this quantity closely related to Planck's constant \cite{Bekenstein}. Hence, it is natural to ponder whether there is a fundamental limit to the accuracy attainable by star sensors, given constraints such as the \emph{volume} in space it occupies, the length of \emph{time} available for observations, and the distribution and brightness of stars around it.

An attitude sensor is any instrument used aboard spacecraft to provide data to estimate its orientation in space (attitude). Many spacecraft need to have an accurate knowledge of their attitude in order to accomplish their mission goals (e.g., point cameras/telescopes, communication antennas, thrusters, etc.). In order to do so, they use a variety of attitude sensors, such as sun and horizon sensors, magnetometers, and star sensors (specifically, \emph{star scanners} for spinning spacecraft and \emph{star trackers} for three-axis stabilized spacecraft) \cite{Wertz1978, Eterno1999}. Star trackers (STRs) are among the most accurate attitude sensors available for spacecraft use, by providing absolute triaxial attitude measurements with errors typically in the order of few arc-seconds or less \cite{Eisenman1997, Wang2012}. These sensors are in essence computerized optical cameras with the appropriate software for star extraction, star identification, and attitude determination. Reference \cite{Liebe2002} provides a good explanation on how star trackers work.

The purpose of this study is to present an estimate for the ultimate limits for attitude determination from stars, imposed by fundamental laws of Physics, i.e., limits that cannot be overcome by technology improvements, for our location in our galaxy. These estimates are useful as a basis for assessing real star sensors as to their potential for improvements through technology advancements. We will not discuss in this work practical limitations faced by existing, real world star sensors, such as readout noise, non-ideal point spread function (PSF) in centroiding, and distortions introduced by the optics, since these limitations have already been well covered by existing literature \cite{Enright2012, Sun2013, Zakharov2013}. Even though our discussion will focus on an idealized star tracker, the limits derived here are also applicable to star scanners and any other possible future type of star sensor, regardless of the technology employed.

This work is organized as follows: Section \ref{sec_Methodology} describes the methodology used, Section \ref{sec_Results} presents and discusses the results, Section \ref{sec_Future} describes how the model adopted here could be improved, and Section \ref{sec_Conclusion} concludes this paper.

\section{Methodology and Model Description}\label{sec_Methodology}

\begin{figure}[htb]
    \centering\includegraphics[width=0.5\textwidth]{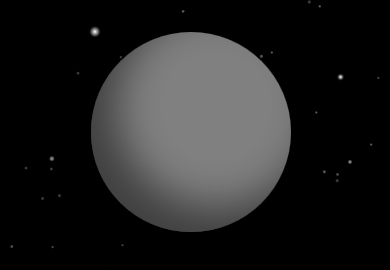}
    \caption{Ideal star tracker (STR) model with stars on the background. To aid visualization, unrealistically represented by a gray sphere here.}
    \label{fig_STRModel}
\end{figure}

The star sensor model analyzed in this work is an ideal \emph{spherical star tracker}, capable of measuring the direction and energy of every photon incident on its surface. This ideal STR is able to observe stars from any direction, that is, it has a field of view of $4\pi\ \rm sr$. The knowledge on the incoming direction of photons in this model is limited only by diffraction at the star tracker aperture, assumed to be circular with the same radius of the STR itself. In other words, it is assumed that the STR aperture is given by the projection of the STR body on a plane perpendicular to the direction of incoming photons. Fig. \ref{fig_STRModel} provides a sketch of the STR model adopted in this work. In this model, the accuracy of the centroids of each star is limited only by diffraction and shot noise. These effects depend only on the STR aperture, stellar spectra, and integration time (exposure time). This ideal STR is completely black, as it absorbs every photon impinging on it. Sections \ref{subsec_Basic} and \ref{subsec_simpl} provide more details on the assumptions adopted in this model.

\subsection{Basic assumptions}\label{subsec_Basic}

The following basic assumptions were made:
\begin{enumerate}
\item \label{Assumption_SphericalSTR} the star tracker has a spherical shape with a diameter $D$;
\item \label{Assumption_allPhotons} it is able to detect \emph{every} photon of stellar origin impinging on its surface;
\item \label{Assumption_photonKnowledge} it is capable of registering the incoming direction and energy of every detected photon with an accuracy limited only by Heisenberg's Uncertainty Principle;
\item \label{Assumption_timeConstraint} only photons detected during a period of length $t$ -- the exposure or integration time -- are considered for attitude determination;
\item \label{Assumption_temperature} it is at absolute zero temperature;
\item \label{Assumption_idealCatalog} the coordinates of the stars in an inertial reference frame with origin in the star tracker are known with absolute precision;
\item \label{Assumption_idealAlg} an unbiased optimal estimator is used to determine the star sensor attitude, and computations are performed with infinite precision;
\item \label{Assumption_noExternalMeas} measurements obtained with this ideal STR are not merged with external measurements; \end{enumerate}

Assumption \ref{Assumption_allPhotons} implies that the STR field of view is $4\pi\ \rm sr$, in other words, it is capable of observing the whole celestial sphere simultaneously, a fact that coupled to its spherical shape, implies that the accuracy of this ideal STR \emph{does not depend on its attitude}.

Assumption \ref{Assumption_photonKnowledge} and the fact that \emph{every} photon is detected implies that the optics are ideal: $100 \% $ transmission, with no \emph{defocusing} and \emph{blurring}, except for the blurring dictated by diffraction.

Assumption \ref{Assumption_temperature} means there is no noise of thermal origin within the star tracker.

Assumptions \ref{Assumption_timeConstraint} and \ref{Assumption_noExternalMeas} limit the number of photons that will be observed by the ideal STR. If exposure time were not constrained, it would be possible to get attitude measurement uncertainty as low as desired, just by increasing the exposure time. In addition, this model does not consider the possibility of combining current observations with previous observations to improve accuracy. Assumption \ref{Assumption_timeConstraint} also implies that the STR is able to measure just photons and no other particles\footnote{The only other particle known to science that could, perhaps, convey better the positions of stars are neutrinos emitted at their core. However, these particles interact so weakly with ordinary matter that their detection in star sensors is currently impossible and may never become a reality\cite{Jayawardhana}.}.

Assumption \ref{Assumption_idealCatalog} implies that the star catalog is perfect and that all corrections needed to bring the coordinates, brightness and colors from the star catalog reference frame origin to the STR location (corrections for stellar aberration, parallax, and redshift/blueshift) are performed with no errors.

Assumption \ref{Assumption_noExternalMeas} expresses the goal of obtaining a lower bound on attitude error for a single star tracker used in isolation. If measurements from multiple sensors were permitted to be merged, a significant improvement in attitude measurement accuracy would become possible. For example, by interferometrically combining measurements from a small number of STRs mounted in a rigid structure and separated by a distance much greater than their diameters, it would be possible to improve attitude determination by many orders of magnitude in comparison to the theoretical estimate presented in this work, with attitude uncertainty being roughly inversely proportional to the distance between them \cite{Baldwin, Molinder}.

\subsection{Simplifying assumptions}\label{subsec_simpl}

In addition to the previous assumptions, to make this study feasible, the following additional assumptions were also made:
\begin{enumerate}
\item \label{SimpAssum_OnlyStars} the whole Universe is assumed to be composed only by stars;
\item \label{SimpAssum_PointSource} stars are considered as polychromatic \textbf{point sources} of light;
\item \label{SimpAssum_BlackBody} stellar spectra are approximated by the \textbf{spectra of black bodies} that best match the cataloged star intensity given by star catalogs adopted here\footnote{Ideally, the actual spectra of stars should be used, at least for the brightest stars, something to be attempted in future works. Section \ref{subsec_AdequacyBB} discusses the adequacy of this approximation.};
\item \label{SimpAssum_NoSolarSystemObj} all Solar System bodies (including the Sun) are \textbf{disregarded};
\item \label{SimpAssum_NoProperMotion} stellar proper motion is \textbf{disregarded};
\item \label{SimpAssum_NoSTRRotation} the star tracker is \textbf{not rotating};
\item \label{SimpAssum_photonOrigin} it is assumed that each detected photon can be univocally associated with the star from where it originated;
\item \label{SimpAssum_StarsAtInf} stars are considered to be at an infinite distance.
\end{enumerate}

Simplifying assumption \ref{SimpAssum_OnlyStars} means that we are not considering as additional sources of attitude information extended bodies, such as interstellar clouds, given that these sources are difficult to precisely model and would hardly significantly increase our attitude knowledge. However, it does not necessarily mean that all non-stellar pointlike sources will be excluded from analysis. This means simply that any non-stellar pointlike source present in star catalogs, such as some quasars and some distant galaxies, will be treated as if they were stars.

Regarding assumption \ref{SimpAssum_NoSTRRotation}, had the star sensor be rotating, but with knowledge of the precise instant each photon were detected and a very accurate knowledge of its own angular velocity vector, it would be possible to compute the incoming direction of every photon in a non-rotating reference frame attached to the star sensor, thus reducing the problem of attitude determination of a spinning star sensor to the problem of attitude determination of a non-rotating star sensor. This makes the limits derived in this work also applicable to star scanners and other possible future types of star sensors. 

Our computations disregard the Sun and other Solar System objects as additional references for attitude determination, since these sources are difficult to model accurately. Also, being the Sun many orders of magnitude closer and brighter than the other stars, from our vantage point in the Universe, it is too bright to be directly observed by most, if not all, star sensors. However, an attitude sensor in a distant future which is able to use and model very accurately the Sun and a neighboring planetary body as additional attitude references, could, perhaps, overcome the estimates on the lower bound of attitude uncertainty computed in this work. This is a topic to be better investigated in the future.

\subsection{Model description}

\begin{figure}[htb]
	\centering\includegraphics[width=0.95\textwidth]{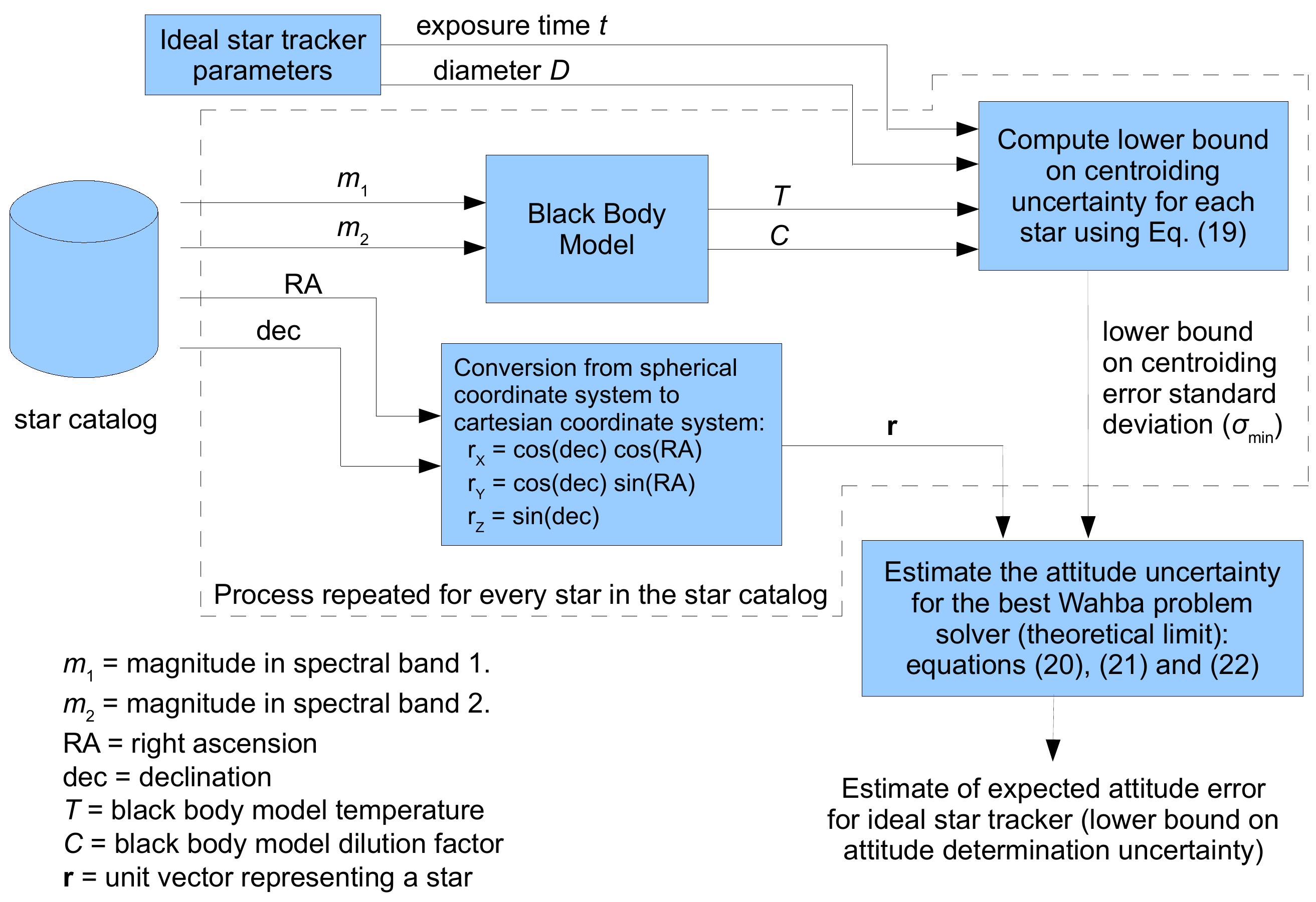}
	\caption{Model for estimating the theoretical lower bound on attitude uncertainty for star sensors.}
	\label{fig_ModelFlowchart}
\end{figure}

Figure \ref{fig_ModelFlowchart} presents a flowchart for the model used in this work. Basically, for each star in the selected star catalog, an estimate for the lower bound on centroiding uncertainty is computed, and these estimates are used together with the unit vectors that represent the stars in the star catalog reference frame to determine the lower bound on attitude determination uncertainty (box at the lower right corner).

Unfortunately, no star catalog is complete. Therefore, any estimate obtained from an existing star catalog will be incomplete, since the missing stars in that star catalog still can contribute to attitude knowledge if they are observed by the star tracker, no matter how far or dim they are. To work around this limitation, we plot the relation of attitude knowledge upper bound with star catalog size for a number of publicly available star catalogs and extrapolate that to the estimated number of stars in our galaxy, plus some margin, to account for extragalactic sources, as described in Section
\ref{subsec_catalogExtrapolation}. In the following sections, a more detailed description of the model used is given.

\subsection{Black body model for stars}\label{subsec_BBmodel}

In the model adopted in this work, the spectrum of each star is considered to be the spectrum of an equivalent spherical black body, diluted by a non-dimensional geometric factor $C$ arising from its distance to the STR. Given that the spectral exitance of a black body is uniquely determined by its temperature, only two parameters are needed in this model to determine the spectral distribution and intensity of the electromagnetic radiation received by the STR from each star, the temperature $T$ and the dilution factor $C$. Mathematically:

\begin{equation} \label{eq_spectral_irradiance}
    E_{e, \lambda, i} (\lambda) = C_i \cdot M_{e, \lambda} (T_i, \lambda)
\end{equation}

\noindent
where:
\begin{itemize}
\item $E_{e, \lambda, i} (\lambda)$ is the spectral irradiance received from star $i$ by a surface located at the same place of the star tracker and perpendicular to incoming rays, evaluated at wavelength $\lambda$;
\item $C_i$ is the geometric dilution factor for star $i$;
\item $T_i$ is the temperature of the black body that represents star $i$;
\item $M_{e,\lambda}(T_i,\lambda)$ is the spectral exitance of the surface of the equivalent black body, at wavelength $\lambda$.
\end{itemize}
In this equation, both $E_{e, \lambda, i} (\lambda)$ and $M_{e,\lambda}(T_i,\lambda)$ are given in unit of power per unit of area and per unit of wavelength (e.g.: W/m$^2$/nm).

To uniquely determine these two parameters ($T$ and $C$) for each star, at least two samples of their flux taken at different wavelengths or at different spectral bands are needed. The following sections describe how $T$ and $C$ are derived for each star from Hipparcos catalog data\footnote{\url{https://www.cosmos.esa.int/web/hipparcos/catalogues}} \cite{ESA}, using the cataloged $m_V$ magnitudes and $B-V$ color indexes. A similar procedure is performed with data from Hipparcos using the $V-I$ color indexes and data from other star catalogs.

\subsection{Black body temperatures from \textit{B -- V} color indexes}\label{subsec_modelBV}

Taking as an example data from the Hipparcos catalog, the spectra of stars is taken as the spectra of black bodies with intensities adjusted so that the integrated spectra over the Johnson's $B$ and $V$ bands \cite{Bessell2005, Bessell2012} match simultaneously the flux at these bands derived from catalog data. To determine equivalent black bodies temperatures for stars in the Hipparcos catalog, an empirical relation is established in this section, linking the $B-V$ color indexes given in the Hipparcos catalog with black body temperatures.

The spectral exitance at wavelength $\lambda$ of a black body at a temperature $T$ can be computed as follows \cite{Budding2007}:

\begin{equation} \label{eq_BB_spectral_exitance}
    M_{e,\lambda}(T,\lambda) = \frac{2\pi h c^2}{\lambda^5} \frac{1}{e ^\frac{hc}{\lambda k T}-1}
\end{equation}

\noindent
where $h$ is the Planck's constant, $c$ is the speed of light in vacuum and $k$ is the Boltzmann constant. The spectral exitance will have units of power per unit area per unit wavelength ($\rm[W \cdot m^{-2} \cdot m^{-1}]$ in SI units). Numerical values of $h$, $c$ and $k$ used in computations were those adopted in the 2019 redefinition of the SI base units\cite{BIPM2019}. 

By integrating the product of the spectral exitance of a black body with the Johnson's $B$ and $V$ bands energy responses it is possible to obtain the black-body fluxes in the $B$ and $V$ bands at its surface. This procedure is described in detail by Bessell in \cite{Bessell2005}, section 1.6 -- Synthetic Photometry:

\begin{equation} \label{eq_BfluxBB}
    \varphi_{BB,B}(T) = \int_{\lambda=0}^\infty M_{e,\lambda}(T,\lambda) R_B(\lambda)\dif{\lambda}
\end{equation}
\begin{equation} \label{eq_VfluxBB}
    \varphi_{BB,V}(T) = \int_{\lambda=0}^\infty M_{e,\lambda}(T,\lambda) R_V(\lambda)\dif{\lambda}
\end{equation}

\noindent
where $\varphi_{BB,B}(T)$ = flux at the surface of a black body at temperature $T$ in the Johnson's $B$ band and $R_B(\lambda)$ = spectral energy response function of the Johnson's $B$ band. Analogously, $\varphi_{BB,V}(T)$ and $R_V(\lambda)$ are quantities related to the Johnson's $V$ band.

\begin{figure}[htb]
	\centering\includegraphics[scale=0.85]{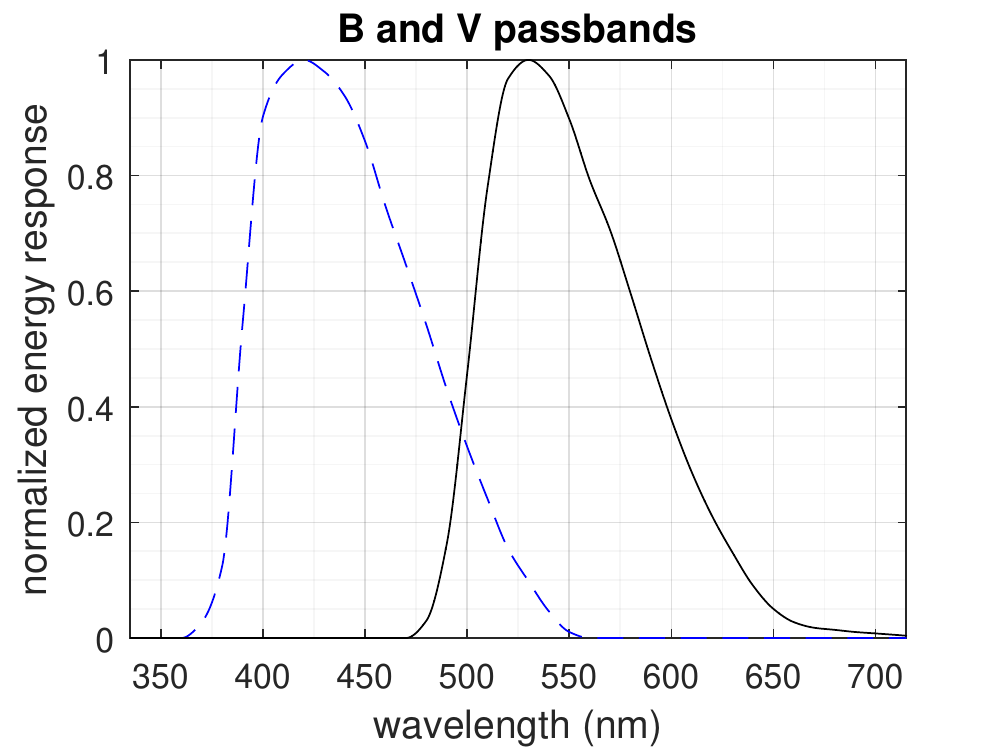}
	\caption{Spectral energy response of the $B$ (blue) and $V$ (visual) bands. The $B$ band is the blue dashed curve to the left.}
	\label{fig_BVresponse}
\end{figure}

The $R_B(\lambda)$ and $R_V(\lambda)$ response functions were obtained by converting the tabulated values recommended by Bessell (Table 1 on page 146 of \cite{Bessell2012}) from normalized photonic responses to normalized energy responses and interpolating the resulting values. The energy response functions adopted in this work are shown in Fig. \ref{fig_BVresponse}.

The conversion from normalized photonic response to normalized energy response was done by multiplying the photonic response by the wavelength and renormalizing the results (Eq. A9 in \cite{Bessell2012}). The explanation for this procedure is given in section A2 in the appendix of \cite{Bessell2012}, on page 153. The method of interpolation used was a ``shape-preserving piecewise cubic interpolation,'' provided by the MATLAB function \verb"interp1" with method ``pchip.'' Computations were performed in MATLAB R2015b with the script \verb"plot_BV_BB_script.m" from the .zip archive which supplements this work\footnote{See ``Supplementary Materials'' on page \pageref{sec_Suppl}.}.

From the fluxes in the $B$ and $V$ bands, the magnitudes in these bands can be computed:

\begin{equation} \label{eq_BmagdefBB}
    m_{BB,B}(T) = -2.5\ \log_{10}(\varphi_{BB,B}(T) / \varphi_{REF,B})
\end{equation}
\begin{equation} \label{eq_VmagdefBB}
    m_{BB,V}(T) = -2.5\ \log_{10}(\varphi_{BB,V}(T) / \varphi_{REF,V})
\end{equation}

These equations give the apparent magnitudes in the $B$ and $V$ spectral bands of a spherical black body for an observer situated just above its surface looking down towards its center. $\varphi_{REF,B}$ and $\varphi_{REF,V}$ are the reference fluxes that define the zero points of the magnitude scales in these bands, having being obtained by numerically integrating the spectrum of Vega ($\alpha$-Lyr) multiplied by the band responses, and adjusting their values such that the computed $B$ and $V$ magnitudes of Vega matches those in the star catalog ($m_{\rm Vega,B} = 0.029$ and $m_{\rm Vega,V} = 0.030$ in Hipparcos). Mathematically:

\begin{equation} \label{eq_Bfluxref}
    \varphi_{REF,B} = 10^{0.4 m_{Vega,B}} \int_{\lambda=0}^{\infty} E_{Vega}(\lambda)R_B(\lambda)\dif{\lambda}
\end{equation}
\begin{equation} \label{eq_Vfluxref}
    \varphi_{REF,V} = 10^{0.4 m_{Vega,V}} \int_{\lambda=0}^{\infty} E_{Vega}(\lambda)R_V(\lambda)\dif{\lambda}
\end{equation}

\noindent
where  $E_{Vega}(\lambda)$ is the spectral irradiance from Vega measured at the top of Earth's atmosphere. The spectrum of Vega used in equations \ref{eq_Bfluxref} and \ref{eq_Vfluxref} was obtained from file \verb"alpha_lyr_stis_008.fits" from the CALSPEC database\footnote{\url{http://www.stsci.edu/hst/instrumentation/reference-data-for-calibration-and-tools/astronomical-catalogs/calspec}.} \cite{Bohlin2014}.

\begin{figure}[htb]
    \centering\includegraphics[scale=0.85]{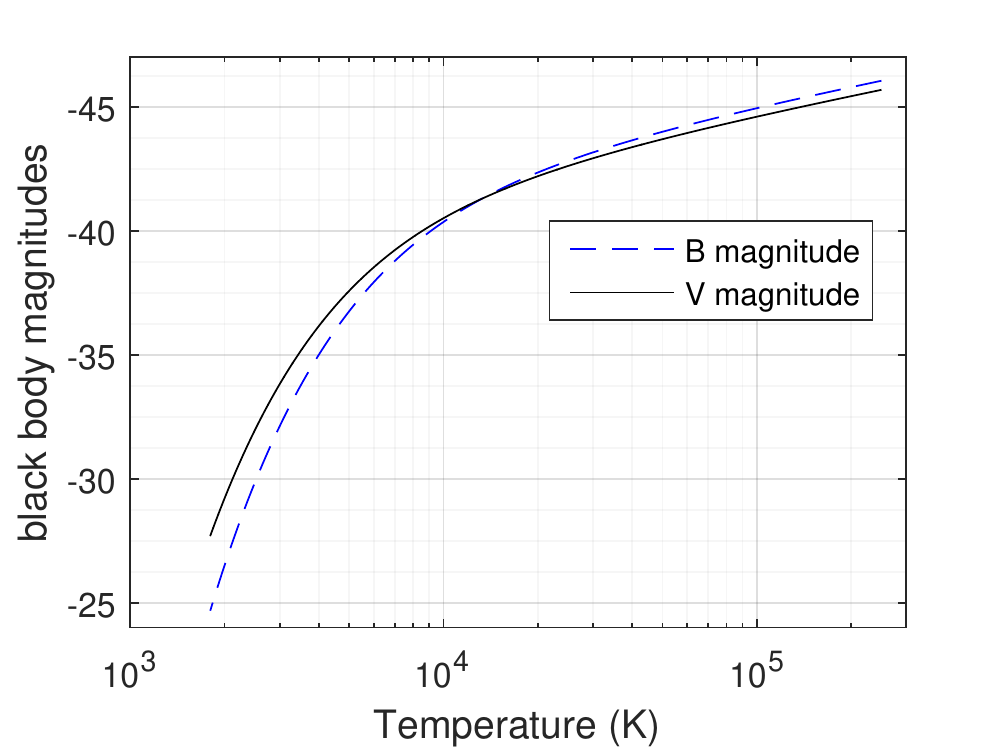}
    \caption{Apparent magnitudes of a black body for an observer lying on its surface and looking down towards its center versus black body temperature, in the Johnson-Morgan $B$ and $V$ bands.
    Note that the vertical axis of this plot is reversed, with more negative magnitudes (brighter black-bodies) at the top.
    }
	\label{fig_BVmags}
\end{figure}

Figure \ref{fig_BVmags} shows the apparent magnitudes of black bodies versus temperature in the Johnson's $B$ and $V$ bands for an observer located at their surface. In this plot, brighter sources (more negative magnitudes) are at the top. Note that the magnitude scale used in astronomy is reversed, with smaller magnitudes meaning brighter sources. The magnitudes are said to be apparent because they depend on the observer location, contrasting to stellar absolute magnitudes which are magnitudes of a star as seen from a standardized distance \cite{Zeilik1998}.

\begin{figure}[htb]
	\centering
    \includegraphics[width=0.42\textwidth]{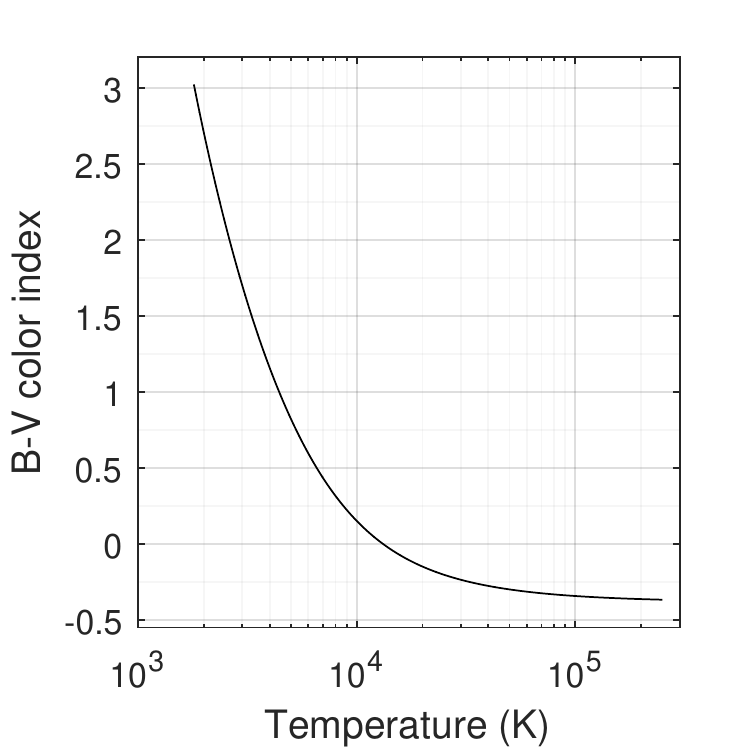}
	\includegraphics[width=0.42\textwidth]{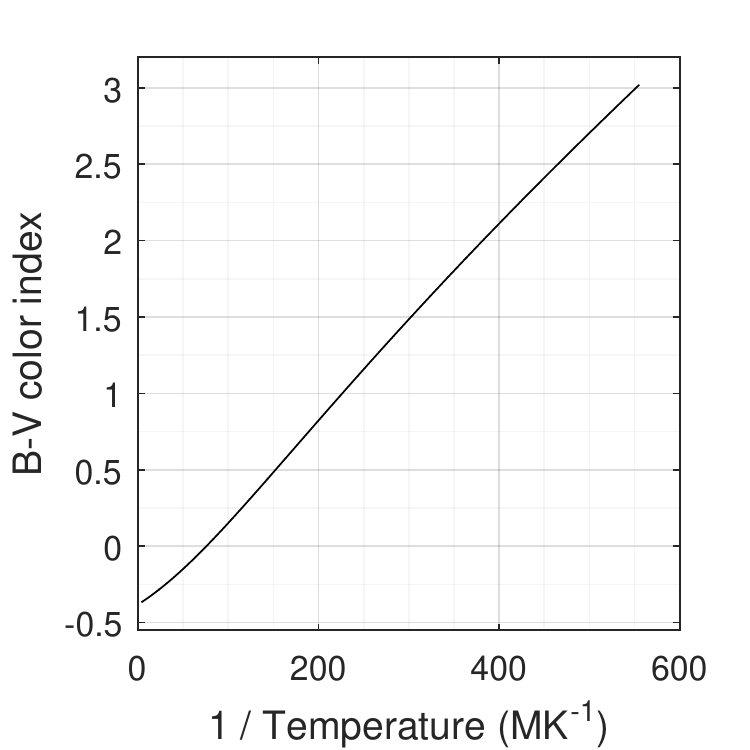}
	\caption{Relation between temperature (or its reciprocal) with $B-V$ color index for black bodies.}
	\label{fig_BVTemp}
\end{figure}

The difference between the $B$ and $V$ magnitudes of a celestial body is its $B-V$ color index. Figure~\ref{fig_BVTemp} presents the relation between the $B-V$ color index and temperature for black-bodies. The plot to the right relates the $B-V$ color index with the multiplicative inverse of its temperature. Note that this curve is much more linear than the direct relation between temperature and $B-V$ color index. Therefore, to get equivalent black-body temperatures for stars in the catalog, we use the 1/$T$ versus $B-V$ curve for interpolation. To avoid temperature estimates with large errors from appearing, the $B-V$ color indexes in the Hipparcos catalog are clamped into the interval $[-0.2357, +2.7028]$ before conversion. These limits correspond to black-body temperatures of $30,\!000$ K and $2,\!000$ K, respectively. Most stars have effective temperatures in that range.

It should be noted that for many stars, the temperature $T$ used in our model will not be equal to the effective temperature\footnote{Effective temperature is the temperature that a black body with the same physical size of a star should have to irradiate the same amount of electromagnetic radiation as the star.} of the star but will usually be smaller. This is caused by interstellar reddening -- selective absorption by dust in the intervening light path from that star to the star tracker. Likewise, the constant $C$ will also have a different value.

\subsection[Determination of the geometric dilution factor]{Determination of the geometric dilution factor $C$ from Hipparcos data}\label{subsec_dilutionFactor}

From temperature $T$, the equivalent black body's visual magnitude at its surface ($m_{BB,V,surface}$) is determined by interpolating the solid black curve in Fig. \ref{fig_BVmags}. The dilution factor $C$ is then obtained by comparing this magnitude with the cataloged visual magnitude ($m_V$) in the Hipparcos catalog, using the following equation:

\begin{equation}
    C_i = 10^{0.4 \cdot (m_{BB,V,surface,i} - m_{V,i})}
\end{equation}

\noindent
The geometric dilution factor $C$ will typically be between $10^{-20}$ and $10^{-14}$ for stars in the Hipparcos catalog. In this equation, the subscript $i$ indicates that the values refer to star $i$.

\subsection{Number of photons detected per unit wavelength}\label{sec_numphotons}

This section derives equations for the number of photons that will be detected, per wavelength, by the idealized star tracker used in this model, for a given exposure time $t$ and a given STR diameter $D$, also assumed to be equal to its aperture diameter. The energy of each photon is related to its frequency $\nu$ by the following equation:

\begin{equation}
    E_{ph} = h\nu = \frac{hc}{\lambda}
\end{equation}

\noindent
Dividing the spectral irradiance at the location of the star tracker due to the black-body equivalent of star $i$ (equations (\ref{eq_spectral_irradiance}) and (\ref{eq_BB_spectral_exitance})) by the energy of a photon of wavelength $\lambda$, the following expression for the spectral photon flux density received by the STR from the equivalent of star $i$ is obtained:

\begin{equation}
    \varphi_{ph,\lambda,i} = C_i \frac{2 \pi c}{\lambda^4} \frac{1}{e^ \frac{hc}{\lambda k T_i} - 1}
\end{equation}

\noindent
This flux density has units of photons per unit of time per unit of area per unit of wavelength. Multiplying this by the star tracker's cross section area $A = \pi D^2/4$ and by the integration time $t$ we obtain:

\begin{equation} \label{eq_numphotons}
    n_{ph,\lambda,i} = C_i \cdot t \cdot \frac{\pi^2 D^2 c}{2 \cdot \lambda^4} \frac{1}{e^ \frac{hc}{\lambda k T_i} - 1}
\end{equation}

\noindent
which is the number of photons from star $i$ equivalent being collected by the STR, per unit wavelength.

\subsection{Diffraction and shot noise}

Diffraction and optics blurring set the format of the point spread function (PSF) of stellar image. In an ideal star tracker, there's no optical blurring, except for that set by diffraction. Therefore, for the STR model adopted in this work, the PSF function will be the diffraction pattern given by a circular aperture of diameter $D$ contained in a plane perpendicular to the incoming direction of photons. This diffraction pattern consists of a disk (Airy disk) with a series of concentric rings, being first derived by Airy in 1835 \cite{Airy1835}.

If the description of Nature given by Classical Mechanics were correct, it would be possible, at least in theory, to measure the intensity of the electromagnetic fields at the detector plane with no error, from where the true, error free, direction of the incoming light rays would be obtained. However, the fact that light is discretized in photons leads to the situation where the number of detected photons will be finite, even with an ideal detector. Therefore, instead of precisely defining the intensity of the electromagnetic fields at each point in the detector (as thought by 19th century physicists), the PSF will define the probability density function that a photon coming from a point source at infinity will be detected on a particular location at the detector. Since the number of photons detected will be finite, even for the case of an ideal star tracker, and these photons are detected at random positions, with probabilities given by the PSF, the centroid estimate for each observed star will have a noise. The lower bound for this noise was determined by Lindegren \cite{Lindegren2013}, being discussed in the next section.

\subsection{Lower bound on centroiding error for single stars}

According to Lindegren \cite{Lindegren2013}, Heisenberg's uncertainty principle sets a fundamental limit for centroiding, and this limit assumes the following form for monochromatic light of wavelength $\lambda$:

\begin{equation}\label{eq_Lindegren}
    \sigma_{xc} \geqslant \frac{\lambda}{4 \pi \Delta x \sqrt{N}}
\end{equation}

\noindent
where:
\begin{itemize}
\item $\sigma_{xc}$ = angular centroiding uncertainty along an axis $x$ perpendicular to the direction of incoming photons, in radians;
\item $\Delta x = \sqrt{\int (x - \overline{x})^2\,\dif{S}\,/ \int \dif{S}}$ is the root mean square extension of the star tracker aperture (entrance pupil) along the $x$ axis, being $\overline{x} = \int x \,\dif{S}\,/ \int \dif{S}$ the position in $x$ of the aperture geometric center; and
\item $N$ = number of photons detected.
\end{itemize}

For circular apertures of diameter $D$, $\Delta x = D/4$. Substituting this into Eq. (\ref{eq_Lindegren}) the following expression for the reciprocal of the lower bound of variance of centroiding error (the Fisher information $F$) over a circular aperture of diameter $D$, for monochromatic sources of light, is obtained:

\begin{equation} \label{eq_var_mono}
    \frac{1}{\sigma_{xc}^2} \leqslant \frac{1}{\sigma_{\min}^2} \triangleq F_{N,\textrm{mono}} = \frac{\pi^2 D^2 N}{\lambda ^2}
\end{equation}

Since stars are incoherent sources of light, the detection of a given photon is not correlated with the detection of another photon from the same star. This means that the number of detected photons from a given star will follow a Poisson distribution with parameter $\iota$, being $\iota$ the expected number of detected photons\footnote{We are using the Greek letter $\iota$ instead of the more common $\lambda$ for the Poisson distribution parameter to avoid confusion with $\lambda$ used for wavelength.}. This parameter can be obtained by integrating Eq. (\ref{eq_numphotons}). For large values of $\iota$, the Poisson distribution narrows down in comparison to the value of $\iota$. This means that when the expected number of detected photons is significantly large, the true value of the lower bound of centroiding accuracy will be very close to the value predicted by Eq. (\ref{eq_var_mono}) if we substitute $N$ by $\iota$. Numerical tests have shown, assuming that the centroiding error for exactly $N$ detected photons follows a Gaussian distribution with a standard deviation given by Eq. (\ref{eq_Lindegren}), that the error between the actual centroiding error and the value estimated by Eq. (\ref{eq_var_mono}) using $\iota$ in place of $N$ will be smaller than $23\%$ for $N \geq 1$, $6.4\%$ for $N \geq 10$ and $0.51\%$ for $N \geq 100$. It is true that the actual probability density function for centroiding
error along one axis will not be exactly Gaussian, specially for a low number of detected photons, but a Gaussian distribution provides a good approximation, even when only one photon is detected.

Another consequence of the fact that the detection of a given photon is not correlated with the detection of another photon from the same star is that the centroiding error of a centroid computed using photons in the wavelength interval $[\lambda_1, \lambda_2]$ is independent on the centroiding error using photons in the wavelength interval $[\lambda_3, \lambda_4]$ when these intervals do not overlap ($\lambda_2 < \lambda_3$ or $\lambda_4 < \lambda_1$). Therefore, we can consider each wavelength interval individually and then merge the centroid estimates for each wavelength.

For the discrete case of having $n$ independent unbiased estimates of the same physical variable (e.g., the $x$ coordinate of a star centroid), each having a variance $\sigma_i^2$, the best estimate for that variable is obtained by summing these estimates using the reciprocal of their variances as weights (a procedure sometimes known as inverse variance weighting)\cite{Hartung, JamesYen}. In that case, the variance of this optimal estimate will be given by:

\begin{equation} \label{eq_inverse_variance_weighting}
    \sigma_T^2 = \left(\sum_{i=1}^n \sigma_i^{-2}\right)^{-1}
\end{equation}

\noindent
where $\sigma_T^2$ = total variance in the estimate of a scalar physical variable obtained by merging $n$ independent measurements and $\sigma_i^2$ = variance of each individual measurement $i$. Since the spectra of black bodies is continuous, the following adaptation of Eq. (\ref{eq_inverse_variance_weighting}) is used to compute centroid estimates for black bodies:

\begin{equation}
    \frac{1}{\sigma_{xc}^2} = \int_{\lambda = 0}^\infty \frac{\dif{(\sigma^{-2})}}{\dif{\lambda}} \dif{\lambda}
\end{equation}

\noindent
The contribution from each wavelength to the knowledge of the centroid position can be obtained from Eq.~(\ref{eq_var_mono}) by replacing $N$ with $n_{ph,\lambda,i}\dif{\lambda}$, where $n_{ph,\lambda,i} = \dif{N_{ph,i}}/\dif{\lambda}$ is the derivative with wavelength of the number of photons from star $i$ entering the star tracker aperture within an integration time of $t$, as given by Eq. (\ref{eq_numphotons}) from section \ref{sec_numphotons}.
Hence, for each star $i$, the wavelength derivative of the maximum knowledge physically attainable of its centroid position (derivative of its centroiding Fisher information) is given by:

\begin{equation} \label{eq_centroidFisher}
    \frac{\dif{F}}{\dif{\lambda}} = \frac{\dif{\left(\sigma^{-2}_{\min}\right)}}{\dif{\lambda}} = \frac{\pi^2 D^2}{\lambda^2} \, n_{ph,\lambda}
\end{equation}

\noindent
Here we have dropped the subscript $i$ to simplify notation. Plugging Eq. (\ref{eq_numphotons}) into Eq. (\ref{eq_centroidFisher}) yields:

\begin{equation}
    \frac{\dif{F}}{\dif{\lambda}} = C \cdot t \cdot \frac {\pi^4 D^4 c}{2 \cdot \lambda^6} \frac {1}{e^\frac{hc}{\lambda k T} - 1}
\end{equation}

\noindent
Integrating this equation for $\lambda=0$ to $\infty$ gives $F_i$, the Fisher information for stellar centroid $i$, and its reciprocal $\sigma_{\min}^2$, the minimum variance for the centroid position error in $x$ direction, being $x$ an axis perpendicular to the incoming light rays:

\begin{equation} \label{eq_starCentroidFisherInfo}
    \frac{1}{\sigma_{\min}^2} = F_i = \int_{\lambda=0}^{\infty} \frac{\dif{F_i}}{\dif{\lambda}}\dif{\lambda} = 12 \zeta(5) \pi^4 \cdot \frac{k^5}{h^5 c^4} \cdot D^4 t \cdot C_i T_i^5
\end{equation}

\noindent
where $\zeta(5) = 1.0369277551...$ 
is the Riemann zeta function evaluated at 5. Since the aperture is symmetrical, Eq. (\ref{eq_starCentroidFisherInfo}) gives the minimum centroiding variance for star $i$ ($\sigma_{\min,i}^2$) along any axis perpendicular to the direction of incoming light rays. From this equation, it can be noted that the lower bound of the standard deviation on centroiding error along any axis perpendicular to the true direction of the star is proportional to $D^{-2}$ and $t^{-1/2}$, when the number of detected photons is sufficiently large. This means that the star tracker diameter has a much larger effect in the ultimate centroid accuracy and precision than the exposure time.

\subsection{Estimating the lower bound of attitude error from many stars}

This section follows the formulation given by Markley and Crassidis in \cite{Markley2014}, Section 5.5. This formulation is valid when measurement errors are small, uncorrelated and axially symmetric around the true direction of stars, conditions fulfilled by our model, except for ideal star trackers with very small diameters, much less than 1 mm.

According to equations 5.113 and 5.114 in \cite{Markley2014}, the covariance matrix ($\mtx{P}_{\vartheta \vartheta}$) of the rotation vector error ($\boldsymbol{\delta \vartheta}$) for an optimal attitude estimator is the inverse of the Fisher information matrix $\mtx{F}$:

\begin{equation} \label{eq_covariance}
    \mtx{P}_{\vartheta \vartheta} = \mtx{F}^{-1}
\end{equation}

\noindent
with:

\begin{equation} \label{eq_FisherMatrix}
    \mtx{F} = \sum_{i=1}^N \frac{1}{\sigma_i^2} \left[ \mtx{I}_{3\times 3} - \vect{r}_i\true(\vect{r}_i\true)\T\right]
\end{equation}

\noindent
where $\sigma_i^2$ is the measurement variance associated with star $i$, as defined in \cite{Markley2014}, $\mtx{I}_{3\times 3}$ is a $3\times 3$ identity matrix and $\vect{r}_i\true$ is the true direction of star $i$, represented by a unit vector expressed as a $3\times 1$ column matrix. $\vect{r}_i\true$ is given in some reference frame R and $N$ is the number of identified stars used in attitude computation\footnote{Since the attitude error does not depend on the reference frame chosen, any one can be chosen for analysis. For simplicity, Markley and Crassidis have chosen the body frame B as the reference frame R in their analysis, making the true direction of stars in the body frame $\vect{b}_i\true$ equal to $\vect{r}_i\true$, that's the reason they use $\vect{b}_i\true$ instead of $\vect{r}_i\true$ in their formulation.}.

In the ideal STR model adopted in this work, the measurement variance $\sigma_i^2$ is identical to the lower bound of centroiding error variance $\sigma_{\min,i}^2$ given by Eq. (\ref{eq_starCentroidFisherInfo}). This statement is proved in the Appendix.

\subsection{A compact metric for the attitude error}

Even though the covariance matrix $\mtx{P}_{\vartheta \vartheta}$ provides detailed information about the attitude uncertainty, as it has six independent parameters it has the disadvantage of being hard to visualize. Therefore, to perform comparisons, we use a more compact metric derived from it:

\begin{equation} \label{eq_finalVariance}
    (\bar{\vartheta}_{rms})^2 = E\{\vartheta^2\} = \tr(\mtx{P}_{\vartheta \vartheta})
\end{equation}

\noindent
The trace of the covariance matrix $\mtx{P}_{\vartheta \vartheta}$ gives the variance of the overall attitude error, that is, the sum of the variances of the attitude error around the three defining axes of the reference frame. It is also equal to the square of the limiting value of the root mean square (rms) of the angle theta ($\vartheta$) of the Euler axis/angle parameterization of the attitude error when the number of attitude determinations tends to infinity.

When the STR diameter and exposure time are large enough so that most stars contributing to the Fisher information matrix $F$ have many detected photons, the lower bound of the expected rms value of theta ($\bar{\vartheta}_{rms, min}$) can be computed by the Eqs. (\ref{eq_starCentroidFisherInfo}), (\ref{eq_covariance}), (\ref{eq_FisherMatrix}) and (\ref{eq_finalVariance}). These equations can also be rearranged in the following manner, which makes more explicit the dependence of $\bar{\vartheta}_{rms,min}$ with $D$ and $t$:

\begin{equation} \label{eq_lowerbound}
    \bar{\vartheta}_{rms,min} = G \cdot D^{-2} \cdot t^{-1/2}
\end{equation}

\noindent
with

\begin{equation} \label{eq_G}
    G = \sqrt{\frac{h^5 c^4}{12 \zeta(5)\pi^4 k^5}\cdot \tr{\left[\left(\sum_{i=1}^N C_i T_i^5 \left[ \mtx{I}_{3\times 3} - \vect{r}_i\true(\vect{r}_i\true)\T \right]\right)^{-1}\right]}}
\end{equation}

\noindent
$G$ is a constant that depends only on stellar distribution around the star tracker, stellar brightness and on attenuation of stellar light by the intervening medium.

\section{Discussion and Results}\label{sec_Results}

\subsection{Star catalogs used}\label{subsec_starcat}

The Hipparcos star catalog was initially selected because it was until very recently one of the most accurate star catalogs available for precise attitude work. Therefore we had already all the tools needed to process it. Unfortunately, the Hipparcos star catalog having less than 120,000 stars is too short to give an adequate basis for extrapolation. Therefore it was decided to include data from two larger catalogs, the Tycho-2 \cite{Hog2000, Turon2009} with around 2.5 million stars and 2MASS \cite{Skrutskie2006} with about 470 million objects.

The Hipparcos and Tycho-2 star catalogs give magnitudes in the optical regime (near ultraviolet, visible and near infrared), whereas the 2MASS star catalog gives magnitudes in the near/shortwave infrared bands $J$(1.25 $\mu$m), $H$(1.65 $\mu$m) and $K_s$(2.16 $\mu$m).

\subsection{Adequacy of the black-body approximation}\label{subsec_AdequacyBB}

\begin{figure}[htbp]
    \centering\includegraphics[width=0.45\textwidth]{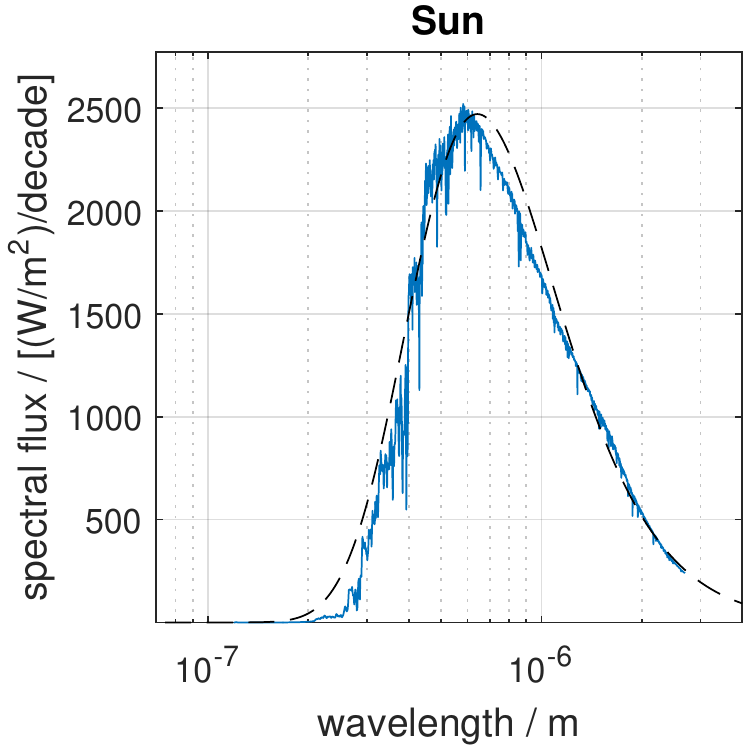} $\qquad$ 
    \includegraphics[width=0.45\textwidth]{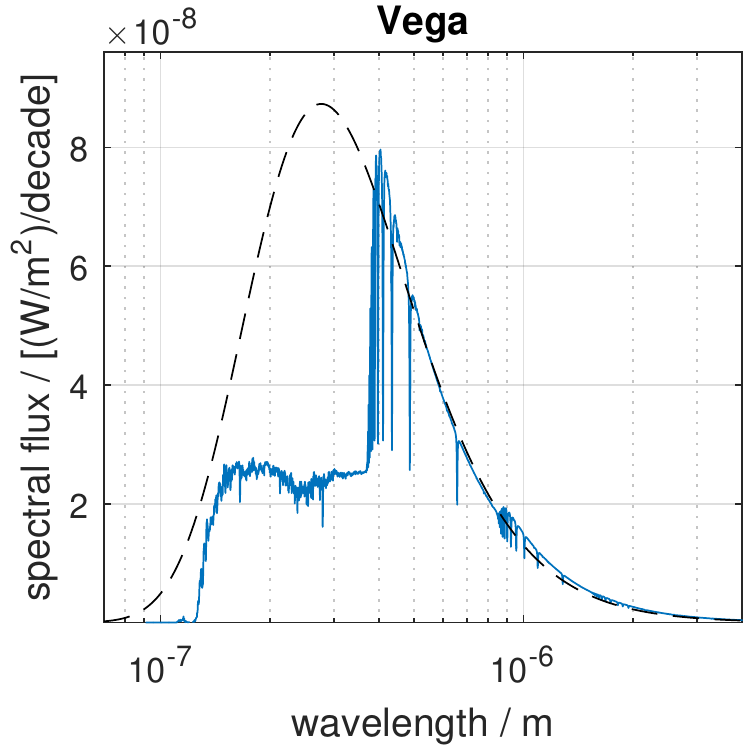}
    \caption{Comparison between the actual spectra for the Sun and Vega	($\alpha$-Lyr) with the spectra of their black-body equivalents derived from their $B-V$ color indexes and $V$ magnitudes with the methodology explained in sections	\ref{subsec_modelBV} and \ref{subsec_dilutionFactor}. Actual spectra represented by continuous line. Dashed lines represent the spectra of equivalent black-bodies.}
	\label{fig_BB_ModelAdequacy}
\end{figure}

In order to check the adequacy of the black-body approximation used in Section \ref{subsec_BBmodel}, we have performed a numerical integration of Eq. (\ref{eq_centroidFisher}) for some selected stars, using their actual spectra. It was observed that, given the color index used, the black-body approximation provides a good fit for some stars, but the fitting is not so good for all of them. Figure \ref{fig_BB_ModelAdequacy} compares the actual spectra of two stars with the spectra of their black-body equivalents, derived from their $B-V$ color indexes and $V$ magnitudes ($m_V$) given in Table \ref{table_SunVega} using the methods described in sections \ref{subsec_modelBV} and \ref{subsec_dilutionFactor}. Spectral fluxes in Figure \ref{fig_BB_ModelAdequacy} are given in power per unit area per frequency (or wavelength) decade.

\begin{table}[htbp]
	\caption{Comparison for some selected stars when $D = 1\,\rm m$ and
	$t = 1\,\rm s$.}
	\label{table_SunVega}
	\centering
	\begin{tabular}{|cc|ccccc|}
	\hline
		\multicolumn{2}{|c|}{Parameter}	&  \multicolumn{5}{c|}{Star}            \\
	name/symbol & unit   & Vega     & 1757132  & Sun*     & KF06T2   & VB8      \\
	\hline
	   spectral type & - & A0V      & A3V      & G2V      & K1.5III  & M7V      \\
	\hline
		$m_V$     & mag  & 0.030    & 11.81    & $-26.75$ & 13.97    & 16.80    \\
		$B-V$     & mag  & $-0.001$ & 0.26     & 0.65     & 1.18     & 2.01     \\
        $ T $     &  K   & 13,231   & 8,580    & 5,711    & 3,951    & 2,613    \\
        $ C $     &  1   & 2.96E-17 & 1.87E-21 & 2.42E-5  & 9.59E-21 & 2.01E-20 \\
$\sigma_{\min,BB}$ & rad & 3.76E-13 & 1.40E-10 & 3.40E-18 & 4.29E-10 & 8.32E-10 \\
    \hline
   		$m_H$     & mag  & $-0.004$ & 11.23    & $-28.24$ & 11.26    & 9.17     \\
		$H-K_s$   & mag  & $-0.005$ & 0.02     & 0.04     & 0.10     & 0.34     \\
		$ T $     &  K   & 10,417   & 8,961    & 8,059    & 6,050    & 3,282    \\
		$ C $     &  1   & 4.80E-17 & 1.94E-21 & 1.41E-5  & 3.72E-21 & 1.05E-19 \\
$\sigma_{\min,BB}$ & rad & 5.37E-13 & 1.23E-10 & 1.88E-18 & 2.37E-10 & 2.06E-10 \\
    \hline
$\sigma_{\min,num}$& rad & 5.50E-13 & 1.75E-10 & 3.61E-18 & 4.67E-10 & 2.92E-10 \\
	\hline
	\multicolumn{7}{l}{\hspace{-6pt}*\, Not used in the results presented in this work, due to its extreme proximity} \\
	\multicolumn{7}{l}{\hspace{-6pt}~~~(see explanation at the end of section \ref{subsec_simpl}).} \\
	\end{tabular}
\end{table}

Table \ref{table_SunVega} also presents a comparison between the lower bound of centroiding error obtained by numerical integration ($\sigma_{\min,num}$, in the last row of the table) and the lower bound of centroiding error $\sigma_{\min,BB}$ obtained from the black-body approximation. To show how $\sigma_{\min,BB}$ can vary depending on the spectral bands used for estimating the equivalent black-bodies, results are presented for two photometric systems: Johnson's UBV and 2MASS JHK$\rm_s$, with the derived black-body parameters ($T$ and $C$) also shown. As can be seen, the error in $\sigma_{\min,BB}$ is typically less than a factor of 2, but sometimes it can be much larger (see for example star VB8).

The magnitudes and color indexes listed in Table \ref{table_SunVega} were computed from spectra downloaded from the CALSPEC library\footnote{\url{http://www.stsci.edu/hst/instrumentation/reference-data-for-calibration-and-tools/astronomical-catalogs/calspec}}. For the UBV system, the zero points that define the origin of the magnitude scales were computed using the method described in section \ref{subsec_modelBV}. For the 2MASS JHK$\rm_s$ system, the zero points were considered to be the zero-magnitude in-band fluxes listed on the third column of Table 2 from Cohen et al. \cite{Cohen2003}.

\subsubsection{Color index limiting values}

As explained in Section \ref{subsec_modelBV}, the color indexes were limited to the interval that corresponds to a temperature range of $2,\!000$ K to $30,\!000$ K. It was observed that, when the upper temperature limit was raised to more than $100,\!000$ K, the Fisher information matrix would be dominated by a few very blue, hot stars where the interpolation from the color index curve versus temperature would give a very high temperature, much higher than their actual temperatures, leading to a significant underestimate of $\bar{\vartheta}_{rms,min}$. In fact, even the $30,\!000$ kelvins upper limit adopted in this work might be too high, resulting that the $\bar{\vartheta}_{rms, min}$ estimated here is probably lower than the actual lower bound of attitude error attainable by star trackers.

The lower limiting temperature of $2,\!000$ K could perhaps be set to a lower value (e.g.: 500 K), in order to better accommodate interstellar absorption and the existence of brown dwarfs. However, it was noted that this lower temperature limit has very little effect in the estimated value of $\bar{\vartheta}_{rms,min}$.

The optimal selection of temperature limits to be adopted for the black-body model will be a subject of a future work, if this model is not abandoned in favor of a more accurate stellar spectra model.

\subsection{Results from catalogs and extrapolation}\label{subsec_catalogExtrapolation}

\begin{figure}[htb]
    \centering\includegraphics[width=0.9\textwidth]{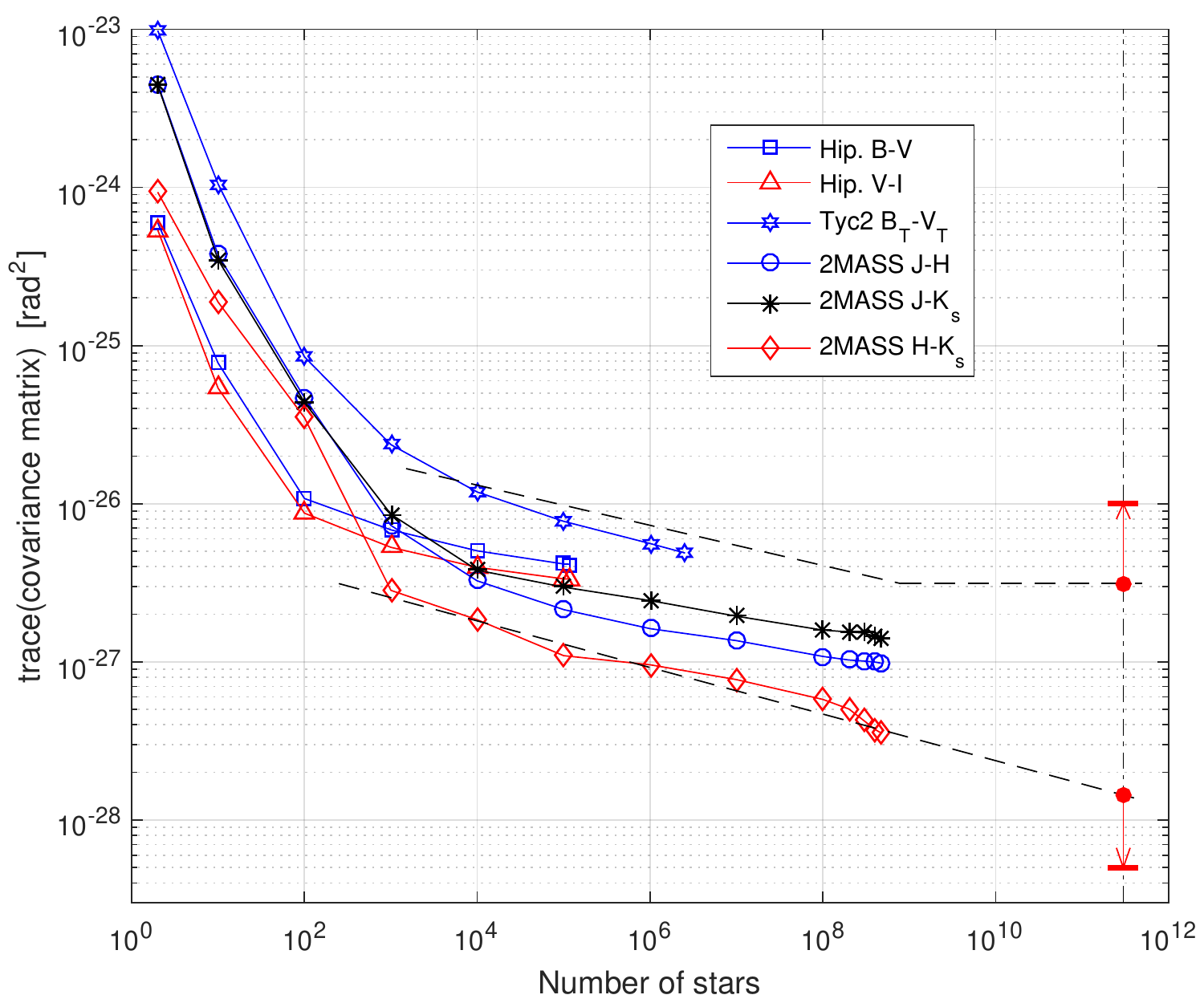}
    \caption{Estimates of $\tr(\mtx{P}_{\vartheta \vartheta})$ for an ideal star sensor with ${D=1}$~m and ${t=1}$~s, obtained from the following star catalogs: Hipparcos, Tycho-2 and 2MASS and their subsets of brightest stars. Extrapolation curves shown as dashed lines. $\tr(\mtx{P}_{\vartheta \vartheta})$ is a measure of the attitude uncertainty (see equations \ref{eq_covariance}-\ref{eq_finalVariance}).}
	\label{fig_extrapolation}
\end{figure}

Some scripts\footnote{See section ``Supplementary Materials'' on page \pageref{sec_Suppl}.} were written to numerically evaluate the lower bound on star tracker attitude error for different star catalogs, different spectral bands and limiting the number of stars used in the computations to the $N$ brightest cataloged stars, with $N$ varying from two stars to the whole star catalog. Figure \ref{fig_extrapolation} shows results obtained with the catalogs described in Section~\ref{subsec_starcat} for  $D = 1$~m and $t = 1$~s. The letter codes B-V, V-I, $\rm B_T$-$\rm V_T$, J-H, J-$\rm K_s$ and H-$\rm K_s$ indicate the spectral bands and catalog used for each curve. These curves form a basis for extrapolation (dashed lines) from where we can obtain an interval that will very likely contain the true value of the parameter $G$ of Equation \ref{eq_G} for our location in the galaxy.

Performing a rough extrapolation, we obtain for N = 300
billion stars\footnote{Many sources give a number between
$10^{11}$ and $4 \cdot 10^{11}$ stars in our galaxy,
with \cite{Guo2009} giving about $3\cdot10^{11}$ stars.
}$^,$\footnote{The word billion can be ambiguous, meaning either $10^{9}$ or $10^{12}$ depending on the language, date and culture \cite{Merr2019,Cambridge2019,RAE2019,GrandDic2019,Lex2019}. In this work we use it as a synonym to $10^9$. A similar ambiguity also exists for the words ``trillion'', ``quadrillion'' and others that follow the same pattern \cite{Lex2019}.
} (the estimated number of stars in our galaxy) $\tr(\mtx{P}_{\vartheta \vartheta}),_{\min} \approx$ $1.4\cdot 10^{-28}$ rad$^2$ for the lower extrapolation curve and $\approx 3.1\cdot 10^{-27}$ rad$^2$ for the upper extrapolation curve. However, there were many approximations made in the model, mainly the assumption of black-body spectra for stars. Therefore, the $\tr(\mtx{P}_{\vartheta \vartheta}),_{\min}$ upper and lower estimates for $D = 1$~m and $t = 1$~s could still be wrong by a factor of 2 or 3. Hence, additional safety factors represented by the red vertical arrows along the $3\cdot10^{11}$ stars dashed dotted line in Figure~\ref{fig_extrapolation} were included. With these safety factors, and considering that the contribution of extragalactic sources is negligible (subsection \ref{subsubsec_Extragalactic}), it should be safe to assume that the true value of $\tr(\mtx{P}_{\vartheta \vartheta}),_{\min}$ lies between $5 \cdot 10^{-29}$ rad$^2$ and $10^{-26}$ rad$^2$ for $D = 1$~m and $t = 1$~s. From this, we conclude that the bounds for the $G$ constant of Eq. (\ref{eq_lowerbound}) are: $7 \cdot 10^{-15}$ rad m$^2$ s$^{1/2} <$ $G$ $< 10^{-13}$ rad m$^2$ s$^{1/2}$, with the upper bound of $10^{-13}$ rad m$^2$ s$^{1/2}$ not being surpassed when $D > 0.1$ m and $t > 0.01$ s.

\subsubsection{Explanation for the upper extrapolation curve in Fig. \ref{fig_extrapolation}}

For $D = 0.1$ m and $t = 0.01$ s (which gives a value of 0.001 m$^2$ s$^{1/2}$ for the combined $D^2 t^{1/2}$ metric used in Fig. \ref{fig_STRcomparison}), it was found that the expected number of detected photons ($\iota$), using values derived from $J-H$ color index in the 2MASS catalog, is positively larger than 1 only for the $8 \cdot 10^8$ brightest stars. For the remaining stars, $\iota$ may be less than 1, meaning that these stars may have a large probability of not being detected. Therefore, conservatively, we ignored these dim stars when building the upper extrapolation curve in Fig.~\ref{fig_extrapolation}, making this curve flat starting at $8 \cdot 10^8$ stars. This makes the attitude uncertainty lower bound ($\bar{\vartheta}_{rms,min}$) obtained from this curve being slightly overestimated.

\subsubsection{Contribution from extragalactic sources}
\label{subsubsec_Extragalactic}

The contribution of all existing extragalactic sources in the known Universe for the attitude accuracy is probably very small (probably less than 10\% of the overall Fisher information). The reason for that is the vast distances between galaxies in comparison to their dimensions. For example, the nearest galaxy about the same size or larger than our galaxy is the Andromeda Galaxy. It's center lies about at a distance of 780 kpc from us \cite{Ribas2005}, which is about 10-20 times the diameter of their disks.

Our galaxy, the Milky Way Galaxy, is orbited by many dwarf galaxies, such as the Small and Large Magellanic Clouds, but the total number of stars in these dwarf galaxies is less than 10\% of the number of stars in our galaxy, therefore their contribution is also negligible.

Considering that the light intensity (and number of detected photons per unit time) falls off with the square of the distance from the source and that the Fisher information contributed by a star is proportional to the number of detected  photons from that source, it is easy to see that the contribution from extragalactic sources will be small.

\subsubsection{Need to consider some stars as extended sources}

The lower bound on attitude uncertainty is so low that future star trackers would probably need to consider some stars as extended bodies and correct the effects of stellar spots (akin to sun spots, but in other stars) in their atmospheres to be able to come close to this theoretical lower bound, something that is unthinkable for current generation star sensors. For example, the star R Doradus, the star with largest apparent diameter after the Sun, has an apparent diameter of 57 $\pm$ 5 mas [(2.76 $\pm$ 0.25)$\cdot 10^{-7}$ rad] \cite{ESO2017}.

\subsection{Comparison with commercial star trackers}

\begin{figure}[htb]
	\centering\includegraphics[width=0.9\textwidth]{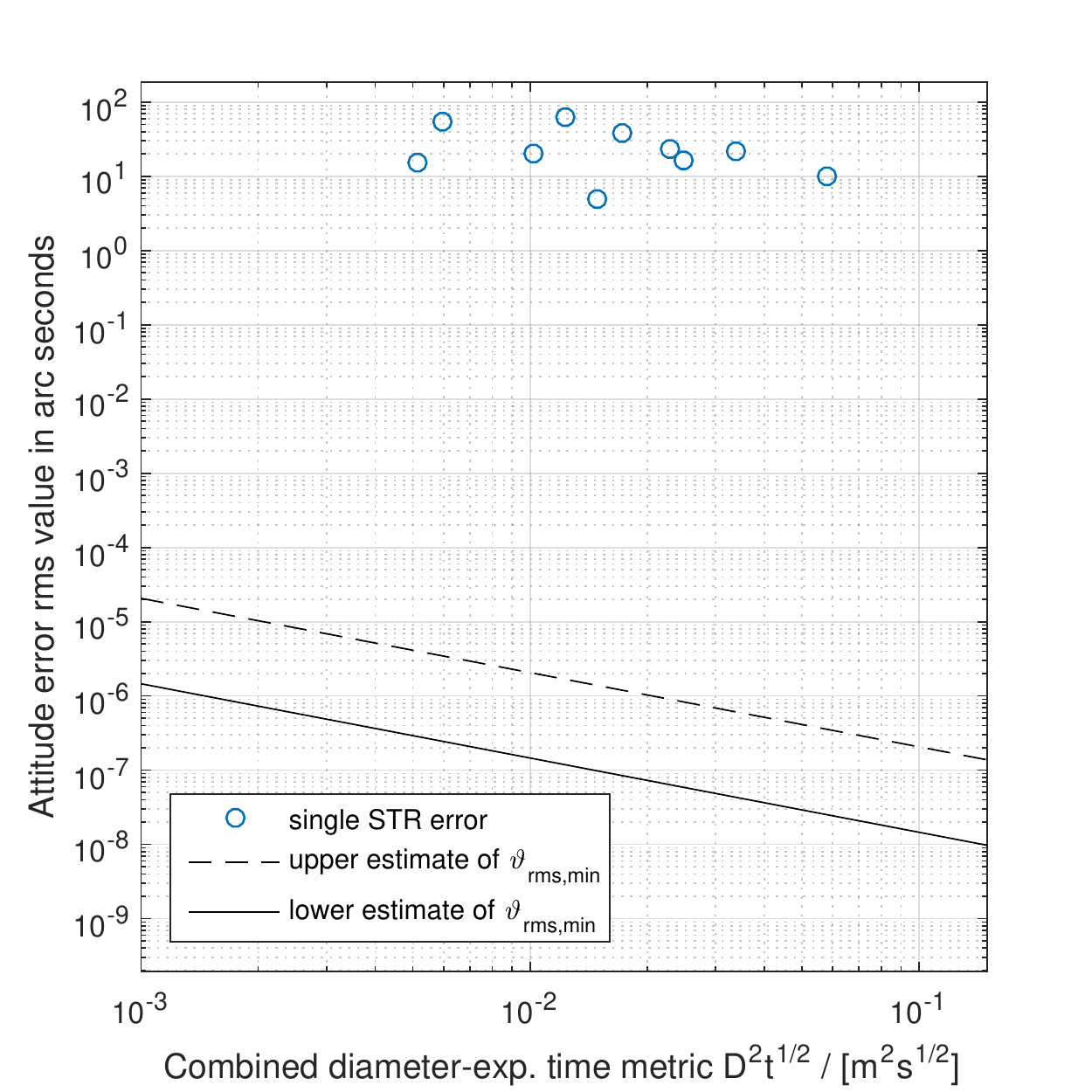}
	\caption{Comparison between some commercial star trackers with the upper and lower limits of the theoretical lower bound on attitude uncertainty.}
	\label{fig_STRcomparison}
\end{figure}

To give a feeling on how much room for improvement there is for future technology developments, Fig. \ref{fig_STRcomparison} compares the reported accuracy of some commercially available star trackers \cite{Astro15, Terma, SED26, Leonardo, VST41M, VST68M, Sinclair, BallCT633, Spacemicro} with the theoretical lower bounds of an equivalent spherical star tracker having approximately the same volume of a sphere that circumscribes the optical head of the star tracker\footnote{the optical head is the box that houses the optics and image sensor. In some models, it includes the whole star tracker with the exception of its baffle. In other models, the processing electronics is in a separate box.}, excluding its baffle\footnote{A baffle is a protective light shade used in star trackers to prevent blinding by the Sun or other bright sources.}.

The comparison is performed in terms of the combined metric $D^2 t^{1/2}$, according to Eq. (\ref{eq_lowerbound}), which makes it possible to compare many different star trackers in a single plot. The solid line at the bottom left of this plot denotes the lower estimate of the lower bound of the attitude error $\bar{\vartheta}_{rms,min}$, derived from the lower curve in Fig. \ref{fig_extrapolation}. The dashed line immediately above this solid line is the upper estimate of the the lower bound of the attitude error $\bar{\vartheta}_{rms,min}$, derived from the upper curve in Fig. \ref{fig_extrapolation}. For $D > 0.1$ m and $t > 0.01$ s the true lower bound of attitude error ($\bar{\vartheta}_{rms,min,true}$) that can be obtained using solely electromagnetic radiation emitted by bodies outside the Solar System should lie between these two curves. No star sensor that satisfies the constraints \ref{Assumption_timeConstraint} and \ref{Assumption_noExternalMeas} stated in Section \ref{subsec_Basic} should be able to surpass $\bar{\vartheta}_{rms,min,true}$ without making use of additional attitude reference sources, such as the Sun and other Solar System objects (excluded from analysis by the simplifying hypothesis \ref{SimpAssum_NoSolarSystemObj} in Section \ref{subsec_simpl}) or artificial attitude references.

We have opted to exclude the baffle in the estimate of $D$, because the sole reason for including a baffle in star trackers is to protect them from being temporarily blinded by stray light coming from the Sun and other bright sources (Earth, Moon and other spacecraft parts), something that would not happen in the absence of Solar System objects. Also, including the baffle would greatly inflate the diameter $D$ of the circumscribing sphere. Unfortunately, star tracker product briefs and other publicly available information only provide their total dimension, including the baffle. Therefore, we had to estimate what would be their sizes without the baffle, by measuring lengths in photos included in the product briefs, from where $D$ could be estimated with an uncertainty probably around $25 \%$. It would be interesting as well to compute the $D^2 t^{1/2}$ metric considering only the optical head aperture, typically much smaller than its dimensions. Unfortunately this information is almost never provided in product briefs. The accuracy used in this plot was derived from the reported (1-$\sigma$) noise equivalent angle or attitude accuracy, using the equation $\vartheta_{rms}^2 = \sigma_x^2 + \sigma_y^2 + \sigma_z^2$, being $\sigma_x = \sigma_y$ the uncertainties around the cross-boresight axes (pitch and yaw angles) and $\sigma_z$ the uncertainty around the boresight axis (roll angle). For single head star trackers, the uncertainty in the roll angle is usually much larger than those in the yaw and pitch angles.

It should be noted that the positions of the commercial star trackers listed in this plot have large errors, being probably wrong by a factor of 2 or 3 in the combined $D^2 t^{1/2}$ metric (horizontal axis) and also by a factor of 2 or 3 in the overall attitude error (vertical axis). The reason for that is the scarcity of information that is available in product briefs and websites. In many cases, to obtain accurate information about star tracker dimensions, exposure time and accuracy it would be necessary to perform lengthy negotiations with star tracker manufacturers.

As can be seen, there's still a lot of room for improvement in star sensors. The theoretical lower bound is around 6 or 7 orders of magnitude lower than what is currently attained by most star trackers.

\section{Future work}\label{sec_Future}

Here are some ideas for future works that could improve the accuracy of the lower bound of attitude error estimates given in this work:
\begin{itemize}
\item{Use more complete star catalogs.} A more accurate estimate on the Cram\'er-Rao bound on attitude accuracy obtainable from stars could be computed using a more complete star catalog, such as the USNO-B star catalog\footnote{\url{http://www.usno.navy.mil/USNO/astrometry/optical-IR-prod/usno-b1.0}} with over 1,000,000,000 
objects\cite{Monet}, or the forthcoming final star catalog from the Gaia astrometric mission\cite{Gaia}\footnote{\url{http://sci.esa.int/gaia/58277-towards-the-final-gaia-catalogue/}}.
\item{Better model for stellar spectra.} Using a more accurate model for the stellar spectra, that takes into account the peculiarities of stellar spectral types and interstellar absorption is something that could be done in the future to improve the accuracy of the results obtained in this work.
\item{Use magnitudes and color index uncertainties provided in star catalogs.} Due to operational limitations we have ignored the uncertainties in magnitude and color indexes provided in star catalogs. An overhaul of the software used to determine the estimates of the lower bound of attitude error is planned for the future. With it, the propagation of uncertainties provided in star catalogs will be included.
\item{More rigorous statistical treatment.} We plan in the future to reevaluate these estimates, using a more rigorous statistical treatment, taking into consideration what happens when the expected number of detected photons from a source is small, less than one. In this scenario, there exists a high probability that no photon from that source will be detected.
\item{Include the Sun and other Solar System bodies in the computation.} It would be interesting to know how much the lower bound on attitude determination uncertainty would be improved if the detected photons from the Sun and other bodies in the Solar System were fully exploited to improve attitude determination, in a system without any technical limitation.
\item{Compute estimates for other parts of the galaxy.} The estimates derived in this work are valid only in the Solar System neighborhood, since they are based on star coordinates, color indexes and magnitudes as seen from the Solar System. It would be interesting in the future to expand this work for other regions of our galaxy. For example, in star dense regions, the ultimate accuracy attainable by star sensors should be better than the ultimate accuracy attainable in our part of the galaxy.
\end{itemize}

\section{Conclusion}\label{sec_Conclusion}

To our best knowledge, Chapter 7 in the doctorate thesis of the first author\cite{Fialho2017} was the first work which provided estimates on the lower bound of attitude uncertainty attainable by star sensors in the Solar System's stellar neighborhood. These derived limits are valid in our stellar neighborhood regardless of the star sensor technology, when no Solar System body can be used to significantly augment attitude determination\footnote{For example, for a spacecraft a few light years from home, in a distant future probably hundreds of years from now, when interstellar travels should become as common as is interplanetary travel today.}.
Being the first work to attempt to derive the numerical values of this limit, it did not aim for much accuracy. This explains the large factor (of about one order of magnitude) between the upper and lower estimates of this lower bound on attitude uncertainty presented in that work\footnote{This article is a thoroughly revised version of that chapter.} and revised here. Nevertheless, these results suffices for the purpose of obtaining an order of magnitude evaluation on how much room for improvement there exists for current state of the art star trackers. It is shown that the accuracy of current star trackers can still improve by about 6 or 7 orders of magnitude before reaching the ultimate limits imposed by laws of Physics and stellar distribution in our stellar neighborhood.

To facilitate verification of the results presented in this work and to foster a culture of open collaboration in the scientific community, the authors make the source code of the routines used to generate results presented in this work available to anyone interested as a free open source software
(see section ``Supplementary Materials'' below).



\vspace{6pt} 


\supplementary{\label{sec_Suppl}The routines used to generate most of the figures and results presented in this work are made available
inside STR\_Limits\_r02.zip under an OSI approved BSD 3 clause copyright license (\url{https://opensource.org/licenses/BSD-3-Clause}).
STR\_Limits\_r02.zip is 512,102 bytes long and its MD5SUM is E8BAE12D0F594440A5F40B86F78A206C.
To assure long term preservation, STR\_Limits\_r02.zip has been published in a public data repository\cite{STRLimits_code} and also in an institutional repository (\url{http://urlib.net/rep/8JMKD3MGP3W34R/3U448R8}).}


\authorcontributions{M.A.A.F developed the software and wrote the bulk of this paper as a part of his doctorate thesis. M.A.A.F and D.M. extensively revised the manuscript.}

\funding{This research was funded by CNPq scholarship grant 207308/2015-2.}


\acknowledgments{The authors would like to thank for the patient review and recommendations given by friends and anonymous reviewers. Special thanks for Dr. Leonel Perondi, who has suggested many improvements to the manuscript. The first author would like also to thank CNPq for the research scholarship 207308/2015-2. Without their financial support, it would be much more difficult to conclude this work.
We also want to express our gratitude to the many people and institutions that in a way or the other contributed and gave support for this research to become a reality, in special, for many people from INPE, CNPq, CAPES, AEB, MCTIC and Texas A\&M University.}




\vspace{6pt} 
\appendixtitles{yes} 
\appendix
\section[Proof that sigma\_i**2 = sigma\_{min,i}**2]{Proof that $\sigma_i^2 = \sigma_{\min,i}^2$} \label{app_sigmaProof}

Being $\sigma_{\min,i}^2$ the lower bound of centroiding error variance for star $i$ along any axis perpendicular to the true direction of that star, given by Eq. (\ref{eq_starCentroidFisherInfo}), and being $\sigma_i^2$ the overall measurement variance associated with that star, used in Eq. (5.114) in \cite{Markley2014}, the goal of this appendix is to prove they are the same for the STR model discussed in this work.

\begin{proof}
\emph{Part 1:}
Let $\vect{b}_i\true$ be a unit vector representing the true position of star $i$ in the star tracker reference frame (body frame) B and $\vect{s}_i\true = [0\ 0\ 1]\T$ the same unit vector in a reference frame S$_i$ where the line joining the star tracker to star $i$ is the $z$-axis of that reference frame\footnote{For each star $i$, there are many different reference frames S$_i$ with this property, but any one of them will suffice for our proof.}. The attitude matrix $\mtx{A}_i$ linking those two frames will satisfy:

\begin{equation} \label{eq_matAdef}
    \vect{s}_i\true = \mtx{A}_i \, \vect{b}_i\true
\end{equation}

Due to measurement errors, the measured direction of star $i$ ($\vect{s}_i$) will differ from its true position $\vect{s}_i\true = [0 \ 0 \ 1]\T$ by $\varDelta \vect{s}_i \equiv$ $\vect{s}_i - \vect{s}_i\true \equiv $ $[\varDelta s_{i,x}\ $ $\varDelta s_{i,y}\ $ $\varDelta s_{i,z}]\T$. Under the assumptions of Section \ref{sec_Methodology} and considering that $\sigma_{\min,i}^2$ gives the lower bound on centroid error \underline{\smash{per axis}}, the expected values of the variances of the $x$ and $y$ components (components in the S$_i$ reference frame) of $\varDelta \vect{s}_i$ will be equal to $\sigma_{\min,i}^2$. In mathematical terms:

\begin{equation}
    E\left \{(\varDelta s_{i,x})^2 \right \} = E\left\{(\varDelta s_{i,y})^2 \right\} = \sigma_{\min,i}^2
\end{equation}

\noindent
with $E\{x\}$ denoting the expected value of a random variable $x$. Since $\vect{s}_i$ is a unit vector very close to $\vect{s}_i\true$ and $\vect{s}_i\true = [0 \ 0 \ 1]\T$, the $z$ component of $\varDelta \vect{s}_i$, $\varDelta s_{i,z}$, will be given by:

\begin{equation}
    \varDelta s_{i,z} \ = \ \sqrt{1 - \Big( (\varDelta s_{i,x})^2 + (\varDelta s_{i,y})^2 \Big)} \,\ - 1 \ \ \approx \ \ - \frac{1}{2} \left ( (\varDelta s_{i,x})^2 + (\varDelta s_{i,y})^2 \right)
\end{equation}

\noindent
As we are retaining only first order terms, like Markley and Crassidis did \cite{Markley2014}, $\varDelta s_{i,z} \approx 0 \; \Rightarrow \; E\{\varDelta s_{i,x} \, \varDelta s_{i,z}\} = E\{\varDelta s_{i,y} \, \varDelta s_{i,z}\} = E\{(\varDelta s_{i,z})^2\} = 0$, in a first order approximation. Given that the star tracker aperture is circular and contained in a plane perpendicular to the direction of incoming rays from star $i$, from symmetry considerations we also have $E\{\varDelta s_{i,x}\, \varDelta s_{i,y}\} = 0$, that is, the errors are axially symmetric about the true vectors $\vect{s}_i\true$. Hence, the measurement covariance matrix for star $i$

\begin{equation} \label{eq_SiDef}
    \mtx{S}_i \equiv E\{ \varDelta \vect{s}_i\ \varDelta \vect{s}_i\T\}
\end{equation}

\noindent
will be in a first order approximation:

\begin{equation} \label{eq_SiApprox}
    \mtx{S}_i \approx \begin{bmatrix} \sigma_{\min,i}^2 & 0 & 0 \\ 0 & \sigma_{\min,i}^2 & 0 \\ 0 & 0 & 0\end{bmatrix}
\end{equation}

\noindent
Its trace will be:

\begin{equation} \label{eq_SiTrace}
    \tr \left(\mtx{S}_i\right) \approx 2 \sigma_{\min,i}^2
\end{equation}

\vspace{6 pt}
\noindent \emph{Part two:}
\vspace{3 pt}

Markley and Crassidis define the following measurement covariance matrix for the errors in the measured star direction vectors $\vect{b}_i$ (Eq. (5.104a) in \cite{Markley2014}):

\begin{equation} \label{eq_Rbi}
    \mtx{R}_{b_i} \equiv E \left\{ \varDelta \vect{b}_i \varDelta \vect{b}_i\T \right \}
\end{equation}

Given the assumption that the vector errors are axially symmetric about the true vectors (in our model, this arises from the consideration that the star tracker has a spherical shape) and ignoring the components along the true star directions (components along vectors $\vect{b}_i\true$ = components $\varDelta s_{i,z}$), since these are of higher order than the terms that we retain, this measurement covariance matrix can be expressed as (Eq. (5.107b) in \cite{Markley2014}):

\begin{equation}
    \mtx{R}_{b_i} = \sigma_{b_i}^2 \left[ \mtx{I}_{3\times 3} - \vect{b}_i\true (\vect{b}_i\true)\T\right]
\end{equation}

\noindent
being $\sigma^2_{b_i}$ the variance in the measured vector position for star $i$ and $\mtx{I}_{3\times 3}$ the $3\times 3$ identity matrix.

Considering that the inverse of an attitude matrix is its transpose, from Eq. (\ref{eq_matAdef}) we have that $\vect{b}_i\true = \mtx{A}_i\T \vect{s}_i\true$, hence:

\begin{equation}
    \mtx{R}_{b_i} = \sigma_{b_i}^2 \left[ \mtx{I}_{3\times 3} - \mtx{A}_i\T \vect{s}_i\true \left( \vect{s}_i\true \right)\T \mtx{A}_i\right]
\end{equation}

\noindent
Considering that the trace of a matrix is a linear operator:

\begin{equation}
    \tr( \mtx{R}_{b_i}) = \sigma_{b_i}^2 \left[\tr(\mtx{I}_{3\times 3}) - \tr\!\left(\mtx{A}_i\T \vect{s}_i\true \left(\vect{s}_i\true \right)\T \mtx{A}_i \right)\right]
\end{equation}

\noindent
Using the matrix trace property, $\tr(\mtx{D} \mtx{C}) = \tr(\mtx{C} \mtx{D})$, with $\mtx{D} = \mtx{A}_i\T \vect{s}_i\true $ and $\mtx{C} = \left(\vect{s}_i\true\right)\T \mtx{A}_i$ and considering that the trace of a $3\times 3$ identity matrix is 3:

\begin{align}
    \nonumber
    \tr( \mtx{R}_{b_i}) &= \sigma_{b_i}^2\left[ 3 - \tr\!\Big(\big( (\vect{s}_i\true )\T \mtx{A}_i \big)\big( \mtx{A}_i\T \vect{s}_i\true \big)\Big) \right] \\
    &= \sigma_{b_i}^2 \left[ 3 - \tr\!\Big((\vect{s}_i\true )\T\big(\mtx{A}_i \mtx{A}_i\T \big) \vect{s}_i\true\Big) \right]
\end{align}

\noindent
Since $\mtx{A}_i \mtx{A}_i\T = \mtx{I}_{3\times 3}$ and $\vect{s}_i\true = [0\ 0\ 1]\T$, the trace of $\mtx{R}_{b_i}$ reduces to:

\begin{align}\label{eq_trR_sigmabi}
    \nonumber
    \tr( \mtx{R}_{b_i}) &= \sigma_{b_i}^2\left[ 3 - \tr\!\Big((\vect{s}_i\true )\T \vect{s}_i\true \Big) \right] \\
    &= \sigma_{b_i}^2 \left[ 3 - \tr([1]) \right] = 2 \, \sigma_{b_i}^2
\end{align}

Considering that both the expectation $E\{\cdot\}$ and matrix trace are linear operators and using the identity $\tr(\mtx{D} \mtx{C}) = \tr(\mtx{C} \mtx{D})$, the following result is obtained from Eq. (\ref{eq_Rbi}):

\begin{align}
    \nonumber
    \tr(\mtx{R}_{b_i}) &= \tr\!\big(E\{\varDelta \vect{b}_i \varDelta \vect{b}_i\T\}\big) = E\big\{\!\tr(\varDelta \vect{b}_i \varDelta \vect{b}_i\T)\big\} \\
    &= E\big\{\!\tr(\varDelta \vect{b}_i\T \varDelta \vect{b}_i)\big\}
\end{align}

\noindent
In the same manner that $\vect{b}_i\true = \mtx{A}_i\T \vect{s}_i\true$, we have $\varDelta\vect{b}_i = \mtx{A}_i\T \varDelta\vect{s}_i$. Using the matrix property, $(\mtx{B}\mtx{C})\T = \mtx{C}\T\mtx{B}\T$, we also have $\varDelta\vect{b}_i\T = \varDelta\vect{s}_i\T \mtx{A}_i$. Substituting these into the last equation:

\begin{align}
    \nonumber
    \tr(\mtx{R}_{b_i}) &= E\big\{\!\tr(\varDelta \vect{s}_i\T \mtx{A}_i\mtx{A}_i\T \varDelta \vect{s}_i)\big\} = E\big\{\!\tr(\varDelta \vect{s}_i\T \varDelta \vect{s}_i)\big\} \\
    &= E\big\{\!\tr(\varDelta \vect{s}_i \varDelta \vect{s}_i\T)\big\} = \tr\!\big(\!E\{\varDelta \vect{s}_i \varDelta \vect{s}_i\T\}\big)
\end{align}

\noindent
But, from Eq. (\ref{eq_SiDef}), $E\{\varDelta \vect{s}_i \varDelta \vect{s}_i\T\}$ is the measurement covariance matrix $\mtx{S}_i$ for star $i$, whose first order approximation is given by Eq. (\ref{eq_SiApprox}) and trace by Eq. (\ref{eq_SiTrace}). Therefore:

\begin{equation} \label{eq_trR_variance_min}
    \tr( \mtx{R}_{b_i}) = \tr( \mtx{S}_i) = 2 \sigma_{\min,i}^2
\end{equation}

\noindent
From Eqs. (\ref{eq_trR_sigmabi}) and (\ref{eq_trR_variance_min}) the following is obtained:

\begin{equation}\label{eq_sigmamini_sigmabi}
    \sigma_{\min,i}^2 = \sigma_{b_i}^2
\end{equation}

\noindent
Equation (5.109) in \cite{Markley2014} gives $\sigma_i^2$ as:

\begin{equation}
\sigma_i^2 = \sigma_{b_i}^2 + \sigma_{r_i}^2
\end{equation}

\noindent
where $\sigma_{r_i}^2$ is the variance in the reference vector for star $i$. In other words, $\sigma_{r_i}^2$ gives the uncertainty in the cataloged position of that star. Since it is being assumed that these cataloged position are known with no errors (Item \ref{Assumption_idealCatalog} in Section \ref{subsec_Basic}), $\sigma_{r_i}^2 = 0$, implying that  $\sigma_i^2 = \sigma_{b_i}^2$. Substituting this into Eq. (\ref{eq_sigmamini_sigmabi}) the following is obtained:

\begin{equation}
    \sigma_i^2 = \sigma_{\min,i}^2
\end{equation}

\noindent
completing the proof.
\end{proof}

\reftitle{References}


\begin{thebibliography}{999}
\bibitem{Fialho2017} Fialho, M.A.A. \emph{Improved star identification algorithms and techniques for monochrome and color star trackers}, Doctorate Thesis -- Instituto Nacional de Pesquisas Espaciais, S\~ao Jos\'e dos Campos, 2017. Available online: \url{http://urlib.net/8JMKD3MGP3W34P/3PE49N5} (accessed on 29 January 2019).
\bibitem{Moore} Moore, G.E. Cramming more Components onto Integrated Circuits, \emph{Electronics}, Vol. 38, No. 8, pp. 114-117, April 9, 1965. Republished in \emph{IEEE Solid-State Circuits Society Newsletter}, Vol. 20, issue 3, Sept. 2006.
\bibitem{Ball} Ball, P. Moore for less, \emph{Nature}, 21 April 2005. \doi{10.1038/news050418-14} (accessed on 21 July 2017).
\bibitem{Charles} Charles Jr.,H.L. Miniaturized Electronics, \emph{Johns Hopkins APL Technical Digest}, Vol. 26, No. 4, pp. 402-413, 2005.
\bibitem{Beigel} J. Beigel, It's a Small World And Getting Smaller, \emph{Fierce Electronics}. Dec 20, 2013. Available online: \url{https://www.fierceelectronics.com/components/it-s-a-small-world-and-getting-smaller} (accessed on 19 September 2019).
\bibitem{Bekenstein} Bekenstein, J.D. Energy Cost of Information Transfer, \emph{Phys. Rev. Letters.} Vol. 46, Iss. 10, pp. 623-626, March 9, 1981. \doi{10.1103/PhysRevLett.46.623}.
\bibitem{Wertz1978} Wertz, J.R.(Ed.), \emph{Spacecraft Attitude Determination and Control}. Dordrecht: Reidel, 1978. ISBN: 9-0277-1204-2.
\bibitem{Eterno1999} Eterno, J.S. Attitude Determination and Control. In \emph{Space Mission Analysis and Design}, Third Edition, J. R. Wertz and W. J. Larson, Eds. New York: Springer, 1999, pp. 354-380. ISBN: 978-1-881883-10-4.
\bibitem{Eisenman1997} Eisenman, A.R.; Liebe, C.C.; Joergensen,  J.L. The new generation of autonomous star trackers. In \emph{Proc. SPIE}, Vol. 3221, pp. 524-535, 1997. \doi{10.1117/12.298121}.
\bibitem{Wang2012} Wang, H.; Zhou, W.; Cheng, X.; Lin, H. Image Smearing Modeling and Verification for Strapdown Star Sensor, \emph{Chinese Journal of Aeronautics}, Vol. 25, pp. 115-123, 2012.
\bibitem{Liebe2002} Liebe, C.C. Accuracy Performance of Star Trackers: a Tutorial, \emph{IEEE Transactions on Aerospace and Electronic Systems}, Vol. 38, issue 2, pp. 587-599, Apr. 2002. \doi{10.1109/TAES.2002.1008988}.
\bibitem{Enright2012} Enright, J.; Sinclair, D.; Dzamba,T. The Things You Can't Ignore: Evolving a Sub-Arcsecond Star Tracker, in \emph{Proc. of the AIAA/USU Conference on Small Satellites, Advanced Technologies III, 2012}, Paper SSC12-X-7. Available online: \url{http://digitalcommons.usu.edu/smallsat/2012/all2012/79/} (accessed on 29 January 2019).
\bibitem{Sun2013} Sun, T.; Xing, F.; You, Z. Optical System Error Analysis and Calibration Method of High-Accuracy Star Trackers, \emph{Sensors}, Vol. 13, pp. 4598-4623, 2013. \doi{10.3390/s130404598}.
\bibitem{Zakharov2013} Zakharov, A.I.; Prokhorov, M.E.; Tuchin, M.S.; Zhukov, A.O. Minimum Star Tracker Specifications Required to Achieve a Given Attitude Accuracy, \emph{Astrophysical Bulletin}, Vol. 68, Iss. 4, pp. 481-493, 2013. \doi{10.1134/S199034131304010X}.
\bibitem{Jayawardhana} Jayawardhana, R. \emph{The neutrino hunters: the chase for the ghost particle and the secrets of the universe}. London: Oneworld Publications, 256p., 2014. eISBN 978-1-78074-327-1
\bibitem{Baldwin} Baldwin, J.E.; Haniff, C.A. The application of interferometry to optical astronomical imaging, \emph{Phil. Trans. R. Soc. Lond. A}, Vol. 360, Iss. 1794, pp. 969-986, 2002. \doi{10.1098/rsta.2001.0977}.
\bibitem{Molinder} Molinder, J.I. A Tutorial Introduction to Very Long Baseline Interferometry (VLBI) Using Bandwidth Synthesis, \emph{DSN Progress Report} 42-46, pp. 16-28. May-June 1978.
\bibitem{ESA} ESA - EUROPEAN SPACE AGENCY, \emph{The Hipparcos and Tycho catalogues: Astrometric and photometric star catalogues derived from the ESA Hipparcos Space Astrometry Mission}. Noordwijk, Netherlands: ESA Publications Division, 1997, Series: ESA SP Series Vol. 1200, ISBN: 9290923997.
\bibitem{Bessell2005} Bessell, M.S. Standard Photometric Systems, \emph{Annu. Rev. Astron. Astrophys.}, Vol. 43, pp. 293-336, 2005. \doi{10.1146/annurev.astro.41.082801.100251}.
\bibitem{Bessell2012} Bessell, M.S.; Murphy, S. Spectrophotometric Libraries, Revised Photonic Passbands, and Zero Points for UBVRI, Hipparcos, and Tycho Photometry, \emph{Publications of the Astronomical Society of the Pacific,} Vol. 124, pp. 140-157, February 2012.
\bibitem{Budding2007} Budding, E.; Demircan, O. \emph{Introduction to Astronomical Photometry}, Second Edition, Cambridge: Cambridge University Press, 2007. ISBN 978-0-521-84711-7.
\bibitem{BIPM2019} BIPM - Bureau International des Poids et Mesures. Le syst\`{e}me international d'unit\'{e}s. 2019. ISBN 978-92-822-2272-0. Available online: \url{https://www.bipm.org/utils/common/pdf/si-brochure/SI-Brochure-9.pdf} (accessed on 24 May 2019).
\bibitem{Bohlin2014} Bohlin, R.C.; Gordon, K.D.; Tremblay, P.E. Techniques and Review of Absolute Flux Calibration from the Ultraviolet to the Mid-Infrared, \emph{Publications of the Astronomical Society of the Pacific}, Vol. 126, pp. 711-732, Aug. 2014.
\bibitem{Zeilik1998} Zeilik, M.; Gregory, S.A. \emph{Introduction to Astronomy and Astrophysics}, Fourth edition. Thomson Learning, 1998. ISBN: 0-03-006228-4.
\bibitem{Airy1835} Airy, G.B. On the Diffraction of an Object-glass with Circular Aperture, \emph{Transactions of the Cambridge Philosophical Society}, Vol. 5, pp. 283-291, 1835.
\bibitem{Lindegren2013} Lindegren, L. High-accuracy positioning: astrometry. In: Huber M.C.E. \emph{et al.} (eds.) \emph{Observing Photons in Space}, Vol. 9 of the ISSI Scientific Report Series pp. 299-311. October 2013. ISBN 978-1-4614-7804-1. \doi{10.1007/978-1-4614-7804-1_16}.
\bibitem{Hartung} Hartung, J.; Knapp, G.; Sinha, B.K. \emph{Statistical meta-analysis with applications}. Hoboken, NJ, USA: Wiley, 2008. 256 p. ISBN 978-0-470-29089-7.
\bibitem{JamesYen} Yen, J. Combining information (Lecture notes on Statistics), 2002. Available online: \url{https://www.nist.gov/document-13811} (accessed on 19 May 2017)  
\bibitem{Markley2014} Markley, F.L.; Crassidis, J.L. \emph{Fundamentals of Spacecraft Attitude Determination and Control}. New York: Springer, 2014. ISBN 978-1-4939-0801-1.
\bibitem{Hog2000} Hog, E. \emph{et al.}, The Tycho-2 Catalogue of the 2.5 Million Brightest Stars, \emph{Astronomy \& Astrophysics}, Vol. 355: L27, 2000.
\bibitem{Turon2009} Turon, C. The Tycho-2 Catalogue: a reference for astrometry and photometry. Combining the most modern and the oldest astrometric data, \emph{Astronomy \& Astrophysics} Vol. 500, pp. 587-588, 2009.
\bibitem{Skrutskie2006} Skrutskie, M.F. \emph{et al.}, The Two Micron All Sky Survey (2MASS), \emph{The Astronomical Journal}, Vol. 131, pp. 1163-1183, Feb. 2006.
\bibitem{Cohen2003} Cohen, M.; Wheaton, Wm. A.; Megeath, S. T. Spectral irradiance calibration in the infrared. XIV. The absolute calibration of 2MASS, \emph{The Astronomical Journal} Vol. 126, pp. 1090-1096, Aug. 2003.
\bibitem{Guo2009} Guo, J.; Zhang, F.; Chen, X. Probability distribution of terrestrial planets in habitable zones around host stars, \emph{Astrophys. Space Sci.}, Vol. 323, pp. 367-373, 2009. \doi{10.1007/s10509-009-0081-z}.
\bibitem{Merr2019} ``billion.'' Merriam-Webster.com. 2019. \url{https://www.merriam-webster.com/dictionary/billion} (accessed on 23 September 2019)
\bibitem{Cambridge2019} ``billion.'' Cambridge Dictionary. 2019. \url{https://dictionary.cambridge.org/dictionary/english/billion} (accessed on 23 September 2019)
\bibitem{RAE2019} ``bill\'on''. Diccionario de la lengua espa\~nola. Real Academia Espa\~nola. \url{https://dle.rae.es/?id=5WQzD1r} (accessed on 23 September 2019)
\bibitem{GrandDic2019} ``billion''. Grand dictionannaire terminologique.  Office qu\'eb\'ecois de langue fran\c{c}aise. 2012. \url{http://www.granddictionnaire.com/ficheOqlf.aspx?Id_Fiche=8872290} (accessed on 25 September 2019)
\bibitem{Lex2019}Lexico. \emph{How Many is a Billion?}. Available online: \url{https://www.lexico.com/en/explore/how-many-is-a-billion} (accessed on 25 September 2019)
\bibitem{Ribas2005} Ribas, I. \emph{et al.}, First Determination of the Distance and Fundamental Properties of an Eclipsing Binary in the Andromeda Galaxy, \emph{The Astrophysical Journal}. Vol. 635, L37-L40, Dec. 10, 2005. \doi{10.1086/499161}.
\bibitem{ESO2017} ESO - EUROPEAN SOUTHERN OBSERVATORY, The biggest star in the sky, Mar. 11, 1997. Available online: \url{http://www.eso.org/public/news/eso9706/} (accessed on 2 October 2017).
\bibitem{Astro15} Jenaoptronik, Autonomous Star Sensor ASTRO 15, 2015. Available online: \url{http://www.jena-optronik.de/en/aocs/astro15.html} (accessed on 23 May 2017).
\bibitem{Terma} TERMA A/S Space, Star Tracker HE-5AS, 2012. Datasheet.  Available online: \url{https://www.terma.com/media/101677/star_tracker_he-5as.pdf} (accessed on 23 May 2017).
\bibitem{SED26} Sodern. SED26 Star Tracker, 2013. Available online: \url{http://www.sodern.com/sites/en/ref/SED26_45.html} (accessed on 23 May 2017). 
\bibitem{Leonardo} Leonardo - Societ\`{a} per azioni, Autonomous Star Trackers, 2017. Available online: \url{http://www.leonardocompany.com/en/-/aastr} (accessed on 24 May 2017).
\bibitem{VST41M} Vectronic Aerospace GmbH, Star Tracker VST-41M. Available online: \url{http://www.vectronic-aerospace.com/space-applications/star-sensor/} (accessed on 24 May 2017).
\bibitem{VST68M} Vectronic Aerospace GmbH, Star Tracker VST-68M. Available online: \url{http://www.vectronic-aerospace.com/space-applications/star-tracker-vst-68m/} (accessed on 24 May 2017).
\bibitem{Sinclair} Sinclair Interplanetary, Second Generation Star Tracker (ST-16RT2), 2016. Available online: \url{http://www.sinclairinterplanetary.com/startrackers} (accessed on 24 May 2017).
\bibitem{BallCT633} Ball Aerospace \& Technologies Corp, CT-633 Stellar Attitude Sensor, Boulder, CO, USA, 1998 or 2005. datasheet.
\bibitem{Spacemicro} Space Micro Inc., $\mu$STAR Tracker, 2015. Available online: \url{http://www.spacemicro.com/products/guidance-and-navigation.html} (accessed on 24 May 2017).
\bibitem{Monet} Monet, D. G. \emph{et al.}, The USNO-B Catalog, \emph{The Astronomical Journal}, Vol. 125, pp. 984-993, 2003.
\bibitem{Gaia} Prusti, T. \emph{et al.}, The Gaia mission, \emph{Astronomy \& Astrophysics}, Vol. 595, 36 p., 2016. \doi{10.1051/0004-6361/201629272}.
\bibitem{STRLimits_code} Fialho, M. A. A. 2019. \emph{Theoretical Limits of Star Sensor Accuracy test programs and routines} (version 0.2). Zenodo. \doi{10.5281/zenodo.3462388}.
\end{thebibliography}
\end{document}